\documentclass[conference]{IEEEtran}
\IEEEoverridecommandlockouts
\usepackage{cite}
\usepackage{amsmath,amssymb,amsfonts}
\usepackage{algorithmic}
\usepackage{graphicx}
\usepackage{textcomp}
\usepackage{xcolor}
\usepackage{booktabs}
\usepackage{cleveref}
\usepackage{bbm}
\def\BibTeX{{\rm B\kern-.05em{\sc i\kern-.025em b}\kern-.08em
    T\kern-.1667em\lower.7ex\hbox{E}\kern-.125emX}}
\begin{document}

\title{Interference Mitigation using U-Net Autoencoder based system\\
\thanks{This paper is based upon work supported in part by
the National Science Foundation under Grant ECCS-2029948.}
}

\author{\IEEEauthorblockN{Hiten Prakash Kothari}
\IEEEauthorblockA{\textit{Wireless@VT} \\
\textit{Virginia Tech}\\
Blacksburg, VA \\
hitenkothari@vt.edu}
\and
\IEEEauthorblockN{R. Michael Buehrer}
\IEEEauthorblockA{\textit{Wireless@VT} \\
\textit{Virginia Tech}\\
Blacksburg, VA \\
rbuehrer@vt.edu}
}

\maketitle

\begin{abstract}
This paper proposes a U-Net–based autoencoder framework for mitigating interference in communication signals corrupted by noise and diverse interference sources. The approach targets scenarios involving both signal-plus-noise and signal-plus-interference-plus-noise mixtures, including sinusoidal interferers, LFM chirps, QPSK interferers with different sampling rates, and modulated interference such as QAM. The U-Net architecture leverages multiscale feature extraction and skip connections to preserve fine-grained temporal structure while suppressing interference components. Performance is evaluated using bit error rate and compared against conventional cancellation methods. Results show that the proposed method consistently outperforms traditional techniques in low- and mid-SIR regimes, while remaining competitive at high SIRs. Additional experiments examine the autoencoder’s behavior under model mismatch conditions such as carrier offset and colored noise. The study demonstrates that multiscale neural architectures provide a flexible and effective platform for interference mitigation across a wide range of interference types.
\end{abstract}


\section{Introduction}
Modern wireless communication systems are vital to everyday connectivity, from mobile phones and Wi-Fi networks to satellite and military communications. As the demand for faster and more reliable data transmission grows, so does the complexity of the radio frequency (RF) environment. In many practical scenarios, multiple signals are transmitted over shared frequency bands, which leads to interference, a major challenge that directly impacts the performance of communication systems.

One of the most commonly used single carrier digital modulation schemes in wireless systems is Quadrature Phase Shift Keying (QPSK). QPSK has been used as a representative example for majority of this research for convenience. QPSK is favored for its balance between spectral efficiency and robustness to noise. However, when multiple QPSK signals are transmitted simultaneously, such as in adjacent channels or overlapping time slots, inter-signal interference becomes a significant problem. This interference can distort the signal of interest, causing errors during demodulation and degrading overall system reliability. With the rapid growth of satellite communication networks and the deployment of large constellations in low Earth orbit, the occurrence of interference has increased dramatically. This poses challenges not only for traditional communication systems, where spectral congestion reduces link reliability, but also for radio astronomy, where extremely weak cosmic signals can be masked or corrupted by strong satellite transmissions.

Traditional signal processing methods like filtering, successive interference cancellation (SIC), and adaptive equalization have been widely used to mitigate such interference. While these methods work well under ideal conditions, they often struggle when the interference is nonlinear or time-varying. This becomes even more difficult when the interference and the signal of interest use the same modulation scheme, making them hard to separate.

In recent years, deep learning has shown great potential in learning complex patterns and recovering meaningful signals from noisy or distorted inputs. Unlike conventional algorithms that rely on predefined rules, deep learning models can be trained to automatically discover the underlying structure of the data. This makes them promising tools for challenging signal processing tasks, including interference mitigation.

This paper explores the use of deep learning techniques, specifically convolutional neural networks (CNNs) with U-Net architectures, for separating and recovering a QPSK signal of interest that is corrupted by another interferer signal as well as Additive white gaussian noise (AWGN). The goal is to improve performance in interference-heavy environments and demonstrate how data-driven approaches can outperform traditional techniques in certain scenarios.

\section{Problem Statement}

Interference mitigation plays a pivotal role in enhancing the robustness and reliability of wireless communication systems. The ability to recover signals of interest (SoI) from noise and interference significantly improves system performance and spectral efficiency. Initially, simpler scenarios involving only additive white Gaussian noise (AWGN) are addressed, progressing to more intricate scenarios incorporating both AWGN and scaled interference signals. The signals studied as SoIs are Quadrature Phase Shift Keying (QPSK) waveforms. Complex interference scenarios involve QPSK SoIs corrupted by a variety of interference signals like tone signals, QPSK signals, or LFM chirp interference, with comprehensive performance evaluations across different Signal-to-Interference Ratios (SIR). In all of these interference scenarios a constant AWGN noise has also been embedded to create a case of a practical system. In order to test on more realistic impairments, cases where either the signal of interest or the interference is introduced to a frequency offset or a symbol timing offset are examined. Additionally, denoising and mitigation of Quadrature Amplitude Modulated (QAM) interference were also studied to learn the potential of the models on more complex modulating scheme.

\section{Background}
Deep learning has been increasingly applied to interference mitigation and RF signal denoising, often outperforming classical approaches. The following works highlight progress across automotive radar, wireless communication, IoT, and multi-user networks.

Survey work by Oyedare et al. provides a broad review of deep learning methods for interference suppression, identifying challenges such as generalization and real-time deployment \cite{oyedare_interference_2022}. Lancho et al. introduce the RF Challenge dataset and benchmark, showing that UNet and WaveNet architectures achieve orders-of-magnitude BER improvements over classical filtering and linear estimators, and highlighting the potential of data-driven interference rejection approaches \cite{lancho_rf_2024}. For radio communication denoising, Almazrouei et al. use a convolutional denoising autoencoder for spectrogram-based de-noising across IEEE 802.11 protocols \cite{almazrouei_deep_2019}, while Çağkan et al. adapt the DEMUCS architecture with a U-Net and LSTM bottleneck for RF denoising, achieving strong BER and MSE improvements \cite{yapar_demucs_2024}. 

Successive interference cancellation (SIC) has also been reimagined with deep learning. Luong et al. propose SICNet, a DNN-based alternative to classical SIC, robust to CSI uncertainty and user variations \cite{van_luong_deep_2022}, while Goh et al. extend this to ISICNet for iterative cancellation in NOMA, showing resilience against imperfect CSI \cite{yun_goh_iterative_2024}.

In the field of RFI mitigation for radio astronomy, a plethora of work has been explored using deep learning techniques. Yang et al. propose RFI-Net for FAST that produces fine-grained masks directly from raw data with less training data and faster training, outperforming prior CNN methods on simulated and real observations from FAST and Bleien \cite{yang_deep_2020}. Yosr et al. compare supervised YOLO-v3 object detection with an unsupervised convolutional autoencoder for RFI detection and localization on real data, reporting fast inference and high precision for YOLO and competitive precision for CAE in some cases \cite{ghanney_radio_2020}. Gu et al. introduce a multi-scale convolutional attention U-Net (EMSCA-UNet) that improves precision, recall, F1, and IoU over U-Net, RFI-Net, and R-Net on 40-m Yunnan telescope data by emphasizing scale-adaptive features and attention-weighted maps \cite{gu_radio_2024}. Akeret et al. use a U-Net to classify and mask RFI in 2D time-ordered telescope data, showing accuracy competitive with SumThreshold on HIDE \& SEEK simulations and Bleien data and releasing their code for reproducibility \cite{akeret_radio_2017}.  

Research has also been done on developing end-to-end demodulation architectures for CNN networks. These systems are trained to directly output the bits of interest rather than siganls. Zhao et al. designed an end-to-end CNN demodulator that maps received sequences directly to symbol decisions and reports BER improvements over traditional demodulators without hand-crafted features \cite{zhao_end--end_2021}. Qiang et al. target low-SNR QPSK by cascading denoising and CNN-based demodulation with cross-entropy training, reducing BER substantially at very low SNR, including near -20 dB \cite{qiang_demodulation_2021}. Shevitski et al. combine a convolutional autoencoder with a transformer to perform modulation classification, channel effect inference and correction, and direct baseband demodulation, supported by a new dataset and augmentation pipeline \cite{shevitski_digital_2021}.

Our work differs from these prior studies in several key ways. First, while many deep learning approaches rely on spectrograms or image-like representations, our models operate directly on time-domain IQ baseband signals, aligning closely with practical communication receivers. Second, unlike works limited to noise-only or single interferer conditions, we explicitly address challenging mixed scenarios (e.g., QPSK-on-QPSK with different SPS, QPSK with sinusoidal or LFM interferers, colored noise, and frequency/timing offsets). Finally, our system introduces a novel recommender module that leverages estimated SIR and SPS to dynamically select between U-Net based denoising and successive interference cancellation, offering both flexibility and robustness.  

\section{U-Net based Autoencoder}

\begin{figure}
    \centering
    \includegraphics[width=1\linewidth]{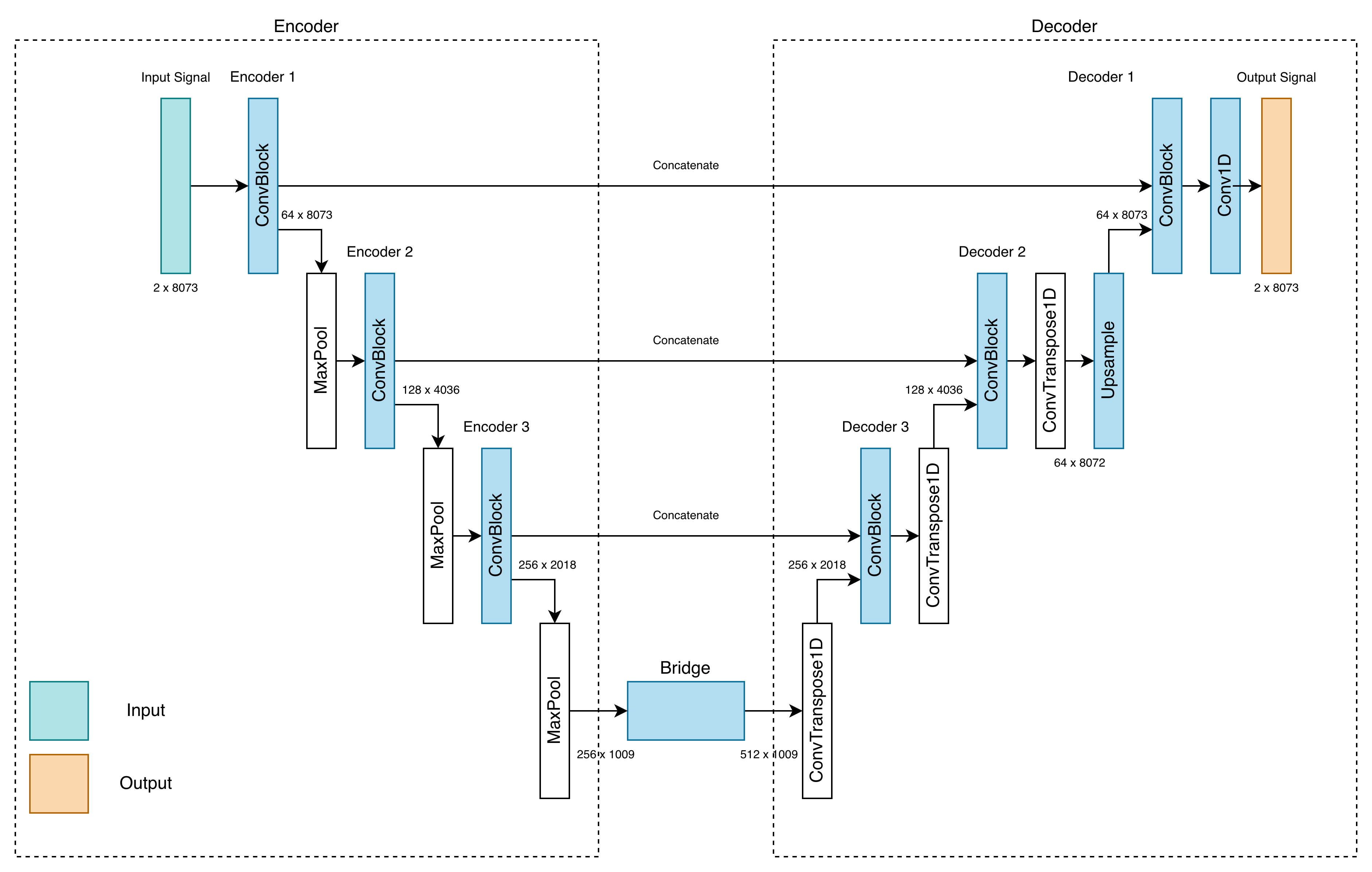}
    \caption{U-Net Autoencoder Architecture}
    \label{fig:u-net_arch}
\end{figure}

\begin{figure}[!htb]
    \centering
    \includegraphics[scale=0.4]{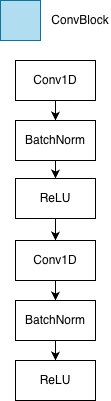}
    \caption{Conv1D Block} 
    \label{fig:u-net_arch2}
\end{figure}

The proposed interference mitigation system in this research employs a convolutional autoencoder with a U-Net-like architecture. The U-Net architecture was originally introduced for biomedical image segmentation \cite{ronneberger_u-net_2015} but has since become a widely adopted framework for signal denoising and sequence modeling tasks. The relationship between CNNs, autoencoders, and U-Nets can be understood as a progression of ideas that build on one another. CNNs are the foundational architecture that apply convolutional filters to learn spatial or temporal patterns in data. Autoencoders extend CNNs by pairing an encoder with a decoder. U-Nets are a specialized form of convolutional autoencoders. They enhance the basic encoder–decoder structure by adding skip connections between corresponding layers of the encoder and decoder. In short, CNNs provide the building blocks, autoencoders organize them into a framework for reconstruction, and U-Nets refine this framework by improving the balance between feature abstraction and detail preservation. The architecture's strength lies in the symmetric encoder–decoder structure, which progressively compresses the input into a compact latent representation and then reconstructs it back to the original resolution. A key feature of U-Net is the presence of skip connections between corresponding encoder and decoder layers. These connections allow the network to retain fine-grained structural details that might otherwise be lost during the encoding process, leading to superior reconstruction quality compared to conventional autoencoders. When adapted to one-dimensional time-series data, such as in-phase and quadrature signals in communication systems, U-Net becomes a powerful tool for denoising and interference mitigation. The autoencoder in this research has been used to recreate a particular type of signal (SoI or Interference) as required by the system for further evaluations. \Cref{fig:u-net_arch} shows the architecture of the U-Net autoencoder used in this research. \Cref{fig:u-net_arch2} goes into detail on the individual stages and the functional blocks of the model.

% Put the auto-encoder architecture here

\subsubsection{Encoder}
The encoder path of U-Net is designed to gradually extract higher-level features from the raw input data. It consists of multiple convolutional blocks, each followed by a downsampling operation such as max pooling. In the context of communication signals, the convolutional layers capture local temporal patterns, such as symbol transitions or interference signatures, while pooling operations reduce the temporal resolution and increase the receptive field. As the signal passes deeper into the encoder, the feature maps become more abstract and descriptive, encoding information about the global structure of the waveform rather than just local variations. This hierarchical extraction of features ensures that the encoder effectively learns both fine-scale and broad-scale representations of the input signal.

\subsubsection{Bottleneck}
At the deepest point of the network lies the bottleneck, which acts as a compressed latent representation of the entire input. In U-Net, the bottleneck consists of convolutional layers that operate on highly reduced feature maps. Despite the reduced resolution, the bottleneck layer encodes global context, capturing interference patterns and signal structures over large time spans. This stage can also incorporate mechanisms such as attention, which selectively emphasizes the most relevant features while suppressing less important ones. The attention block was explored and compared with the non-attention autoencoder and the performance had a slight improvement but at a cost of a more complex model and training/inference time. For interference mitigation tasks, the bottleneck plays a crucial role in separating structured interference from the underlying signal of interest by learning discriminative representations.

\subsubsection{Decoder}
The decoder path mirrors the encoder but in reverse. It begins by upsampling the compressed features back toward the input resolution, typically using transposed convolutions or interpolation methods. Each upsampling step is followed by convolutional blocks that refine the features, guided by skip connections that inject spatially aligned details from the encoder. A skip connection refers to a direct link between corresponding layers in the encoder and decoder paths. Specifically, the feature maps produced at a given level of the encoder are passed forward and concatenated with the feature maps at the matching level of the decoder. These skip connections ensure that temporal structures critical for symbol recovery are preserved during reconstruction. As the signal flows through successive decoder layers, the network progressively rebuilds the waveform, combining global context from the bottleneck with local detail from the encoder. The final output layer maps the reconstructed features back to two channels corresponding to the in-phase and quadrature components of the signal. This reconstruction yields a cleaner version of the signal of interest, suitable for demodulation and bit error rate evaluation.

\section{Training U-Net based Autoencoder}

The parameters used to train the autoencoders are given in \Cref{tab:model_params}. These values were chosen after ablation studies and tuning of hyperparameters. Three layers of Encoder-Decoder blocks were generated for the model to generate sufficient depth to learn the structure of the required signals while keeping the training/inference time quick. For training of the autoencoder, two signals are given as input. One of the signals is the noisy signal which is the mixed signal that we are trying to denoise in these systems. The second signal is the signal of interest that we want the autoencoder to predict. In order to be able to learn hidden features and patterns of the signal of interest, the training was done on a wide range of SNR and SIR.

%put parameter specification table here
\begin{table}[!htb]
    \centering
    \caption{Model Parameters}
    \begin{tabular}{ccc}
        \toprule
        % & \multicolumn{2}{c}{Computation Time}\\
        % \cmidrule(r){2-3}
        Parameters & Value \\
        \midrule
        Batch Size & 256 \\
        Learning Rate & 1e-3 \\
        Epochs & 100 \\
        Examples per SNR/SIR & 5000\\
        Input Size & 8073 x 2 (I and Q)\\
        Number of Parameters & 2,688,194 \\
        Optimizer & Adam \\
        Activation Function & ReLU \\
        Scheduler & ReduceLROnPlateau \\
        Loss Function & MSE Loss \\
        \bottomrule
    \end{tabular}
    \label{tab:model_params}
\end{table}

To ensure that the models were not underfit, several strategies were employed during training. First, the training and validation losses were monitored closely to confirm that both decreased consistently and converged to low, stable values without a large performance gap, indicating effective learning rather than underfitting. The models were trained over a sufficiently large number of epochs, with early stopping used only when convergence was observed. Additionally, the dataset was made sufficiently large so that the model is able to learn the distinct features of signal of interest. Model capacity (number of layers and filters) was also selected empirically, small enough to prevent overfitting, but large enough to capture the nonlinear structure of the signal–interference relationship. Validation accuracy and qualitative inspection of reconstructed or detected signals confirmed that the models successfully learned the underlying signal characteristics instead of underfitting the data.

In this research, two major cases of mixed signals are considered to evaluate the ability of deep learning models for interference suppression. In the first case, the mixed signal consists only of the interference signal combined with noise. This setting provides a baseline to study how well the models can detect and recreate interference. This also works as a simulation for radio astronomy scenario where the interference is strong signals from satellites and other man-mad objects while the cosmic signals of interest are buried within the noise floor. In the second case, an additional signal of interest is introduced alongside the interference and noise, creating a more challenging modern communications scenario. This allows the performance of the models to be assessed under conditions that closely resemble practical wireless environments, where reliable recovery of the signal of interest is essential.

\section{Signal and Noise}\label{sec:s_n}

\subsection{Traditional Approach}

\begin{figure}[!htb]
    \centering
    \includegraphics[width=1\linewidth]{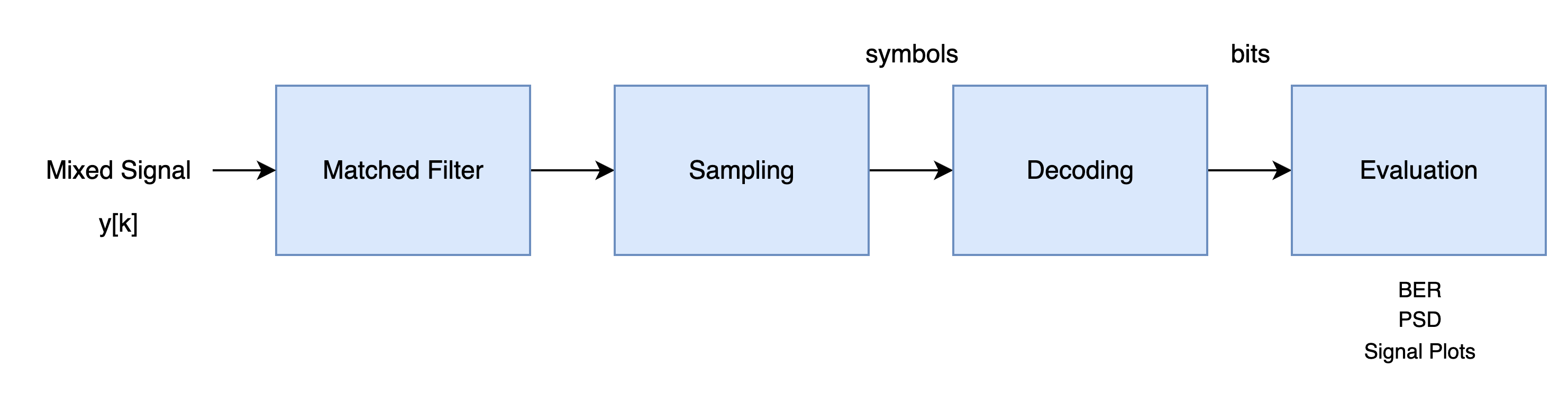}
    \caption{Matched Filter System}
    \label{fig:single_mf_system}
\end{figure}

In the traditional receiver, bit recovery is performed by applying a matched filter to the received baseband signal. The matched filter is constructed as the time-reversed and conjugated version of the transmit pulse shaping filter, in this case the root raised cosine (RRC) filter. The noisy or interference-corrupted signal is directly passed through this matched filter, which maximizes the signal-to-noise ratio at the symbol sampling instants under the assumption of additive white Gaussian noise. After filtering, the output is sampled at integer multiples of the symbol period to obtain estimates of the transmitted symbols. These symbol estimates are then mapped back to their corresponding bit pairs using the QPSK Gray coding scheme. This approach is used as the baseline for comparison with the deep-learning implementation. \Cref{fig:single_mf_system} shows the pipeline used for the conventional approach.

\subsection{System Model}

\begin{figure}[!htb]
    \centering
    \includegraphics[width=1\linewidth]{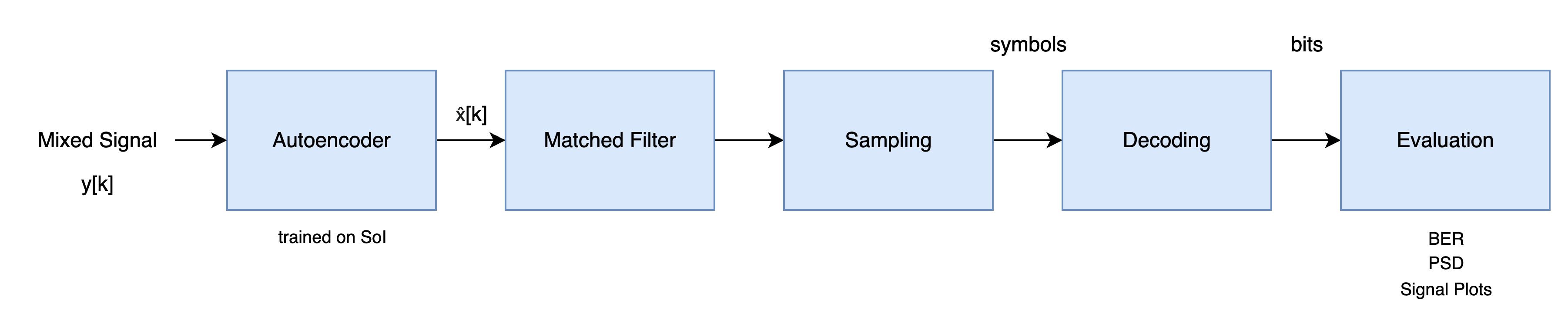}
    \caption{System Model for Signal and Noise Scenarios}
    \label{fig:unet}
\end{figure}

The system model for interference mitigation in the QPSK + noise using autoencoder is shown conceptually in \Cref{fig:unet}. The objective is to reconstruct a clean estimate of the QPSK signal of interest (SoI) from noisy I/Q baseband inputs, enabling reliable symbol detection and bit recovery.

The received noisy signal is expressed as
\[
y[k] = x[k] + n[k],
\]
where $x[k]$ is the transmitted QPSK SoI, and $n[k]$ represents additive noise, which may be AWGN or a colored noise process.  

The denoising pipeline proceeds as follows:
\begin{enumerate}
    \item \textbf{Autoencoder Input:} The noisy baseband sequence $y[k]$ is provided as input to the autoencoder. The input is represented as a two-channel sequence corresponding to the in-phase (I) and quadrature (Q) components.
    
    \item \textbf{Autoencoder Training:} The model is trained in a supervised manner using pairs of noisy signals and their corresponding clean references. The loss function minimizes the difference between the autoencoder output and the clean signal, enabling the network to learn the statistical structure of the QPSK waveform and suppress noise.
    
    \item \textbf{Autoencoder Output:} After training, the autoencoder outputs a denoised estimate of the signal, denoted as $\hat{x}[k]$. This signal is designed to approximate the clean SoI as closely as possible, while attenuating noise components.
    
    \item \textbf{Matched Filtering:} The denoised signal $\hat{x}[k]$ is passed through a matched filter constructed from the known root raised cosine (RRC) pulse shaping filter. The matched filter maximizes the signal-to-noise ratio at the sampling instants and compensates for the pulse shaping applied at the transmitter.
    
    \item \textbf{Symbol Sampling and Demodulation:} The matched filter output is sampled at symbol intervals to recover the transmitted constellation points. These samples are then demodulated into symbol decisions.
    
    \item \textbf{Bit Recovery and BER Calculation:} The detected symbols are mapped back into binary sequences. The recovered bits are compared against the original transmitted bits to compute the bit error rate (BER), which serves as the primary performance metric.
\end{enumerate}

This system model enables a direct comparison between classical matched filtering and the proposed autoencoder-based denoising approach. In the baseline case, matched filtering is applied directly to the noisy input $y[k]$, whereas in the proposed system the matched filter operates on the autoencoder output $\hat{x}[k]$. The difference in BER performance across a range of $E_s/N_0$ values provides quantitative insight into the benefits of learned denoising.

\subsection{QPSK and AWGN}

Traditionally, for this mixed signal matched filtering is considered the optimal linear detection method under AWGN, as it maximizes the signal-to-noise ratio at the decision point. In this study, the proposed autoencoder is trained to reconstruct a cleaner version of the QPSK signal directly from the noisy I/Q input. The performance is evaluated by comparing the reconstructed signals against both the true transmitted signal and the matched filter output. 

\paragraph{Theoretical expression for BER under AWGN:}

The theoretical BER for QPSK signal in an Additive White Gaussian Noise  channel is given by:
\begin{equation}
    P_b = Q\left(\sqrt{2 \frac{E_s}{N_0}}\right)
\end{equation}
where \( \frac{E_s}{N_0} \) is the symbol energy-to-noise power spectral density ratio, and \( Q(\cdot) \) denotes the tail probability of the standard normal distribution:
\begin{equation}
    Q(x) = \frac{1}{\sqrt{2\pi}} \int_x^{\infty} e^{-\frac{t^2}{2}} \, dt
\end{equation}

This expression serves as the theoretical lower bound for BER performance and is commonly used as a benchmark to evaluate the efficiency of practical receiver architectures and learning-based interference mitigation systems.

\begin{figure}[!htb]
    \centering
    \includegraphics[scale=0.5]{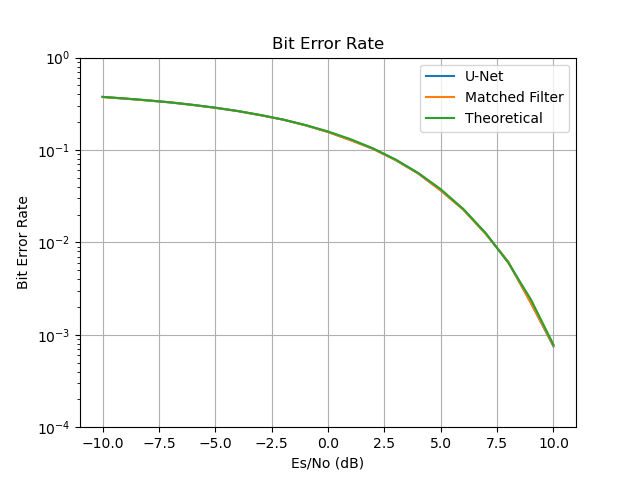}
    \caption{Bit Error Rate across Es/No for QPSK + AWGN} 
    \label{fig:ber_qpsk}
\end{figure}

\begin{figure}[!htb]
    \centering
    \includegraphics[scale=0.5]{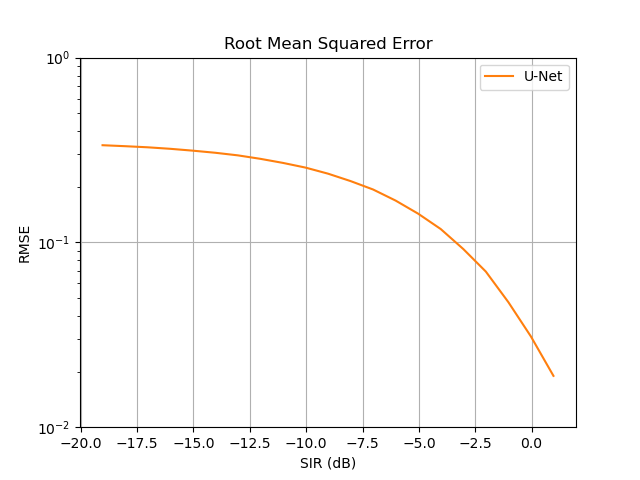}
    \caption{Root Mean Square Error across Es/No for QPSK + AWGN} 
    \label{fig:mse_qpsk}
\end{figure}

As can be observed from the \Cref{fig:ber_qpsk}, the auto-encoder performs on par with the matched-filter and theoretical bit error rate. In the scenario where the signal is corrupted with only AWGN, the autoencoder will not be able to outperform the theoretical best performance which is given by the matched filter. \Cref{fig:mse_qpsk} shows the Root Mean Squared Error (RMSE) of the estimated signal by the autoencoder and the true signal of interest.

\begin{figure}[!htb]
    \centering
    \includegraphics[scale=0.5]{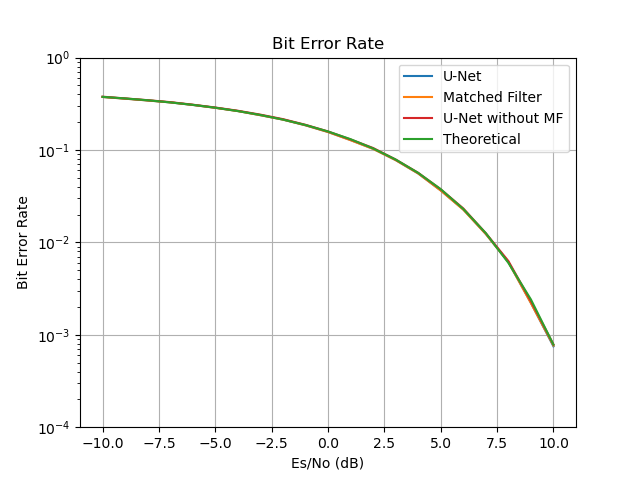}
    \caption{Bit Error Rate without Matched Filter} 
    \label{fig:ber_mf_qpsk}
\end{figure}

\begin{figure}[!htb]
    \centering
    \includegraphics[scale=0.5]{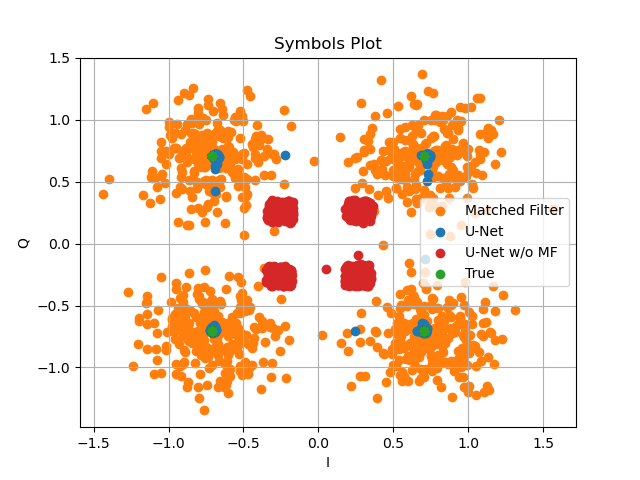}
    \caption{Symbols Plot for QPSK+AWGN} 
    \label{fig:symbol_mf_qpsk}
\end{figure}

\paragraph{Performance without Matched Filter}
The autoencoder being trained to recreate the pulse-shaped signal is able to denoise the noise from the input signal and achieve performance on-par with the Matched Filtered approach. The Matched Filter used after the autoencoder helps in providing a slight improvement in the SNR of the symbols, which also leads to an improvement in BER in some cases. This can be clearly observed in \Cref{fig:ber_mf_qpsk,fig:symbol_mf_qpsk}. The BER of U-Net with and without Matched Filter overlap each other. The symbol plot also showcases that autoencoder is successful at reducing the variance of the signal but has a lower amplitude value. Applying MF after the autoencoder's output helps with a better mapping of QPSK symbols. This was also tested for all the other cases as well. In the cases of impairments, the increase in SNR value after Matched Filter helped in avoiding crossover of symbols to adjacent regions. Therefore, the Matched Filter was kept in the system design pipeline throughout these studies.

\paragraph{End-to-End based U-Net Demodulator}
In this section, an end-to-end deep learning-based demodulator is developed to directly recover the transmitted QPSK bits from the noisy received signal. The end-to-end demodulator is constructed using the same U-Net autoencoder architecture previously employed for denoising, with an additional fully connected layer appended at the end of the decoder to map the extracted features into bit probabilities. A sigmoid activation function follows this layer, producing outputs in the range $[0,1]$, which are then thresholded at 0.5 to obtain binary bit estimates. This design enables the network to jointly learn both denoising and demodulation tasks within a single framework, effectively replacing the conventional matched filter and symbol decision stages. The model is trained on QPSK signals corrupted by additive white Gaussian noise (AWGN) across a range of signal-to-noise ratios, allowing it to learn robust representations that generalize well to varying noise conditions. This model was trained on minimizing the Binary Cross Entropy loss between the true and predicted bits. 

\begin{figure}[!htb]
    \centering
    \includegraphics[scale=0.5]{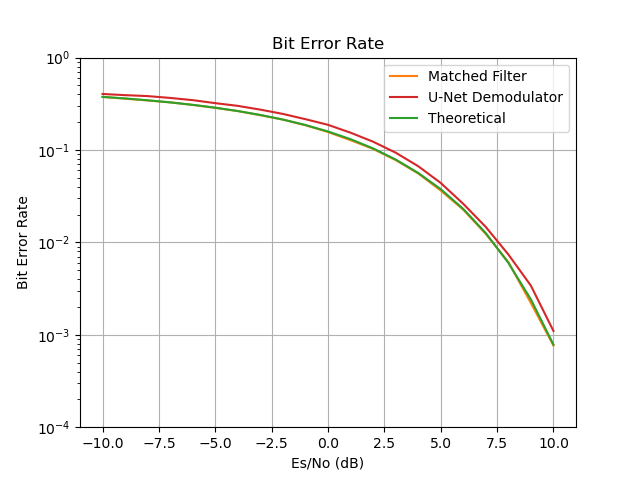}
    \caption{Bit Error Rate across Es/No for QPSK + AWGN using U-Net Demodulator} 
    \label{fig:ber_qpsk_demod}
\end{figure}

\Cref{fig:ber_qpsk_demod} displays the bit error rate of the end-to-end U-Net Demodulator against the traditional demodulating technique and the theoretical baseline. The model's accuracy was slightly worse than the accuracy obtained using the denoising model [\Cref{fig:ber_qpsk}]. This end-to-end approach demonstrates the capability of deep neural networks to jointly perform feature extraction, denoising, and symbol decision within a unified framework.

\subsection{QPSK with Frequency Offset and AWGN}
\begin{figure}[!htb]
    \centering
    \includegraphics[width=1\linewidth]{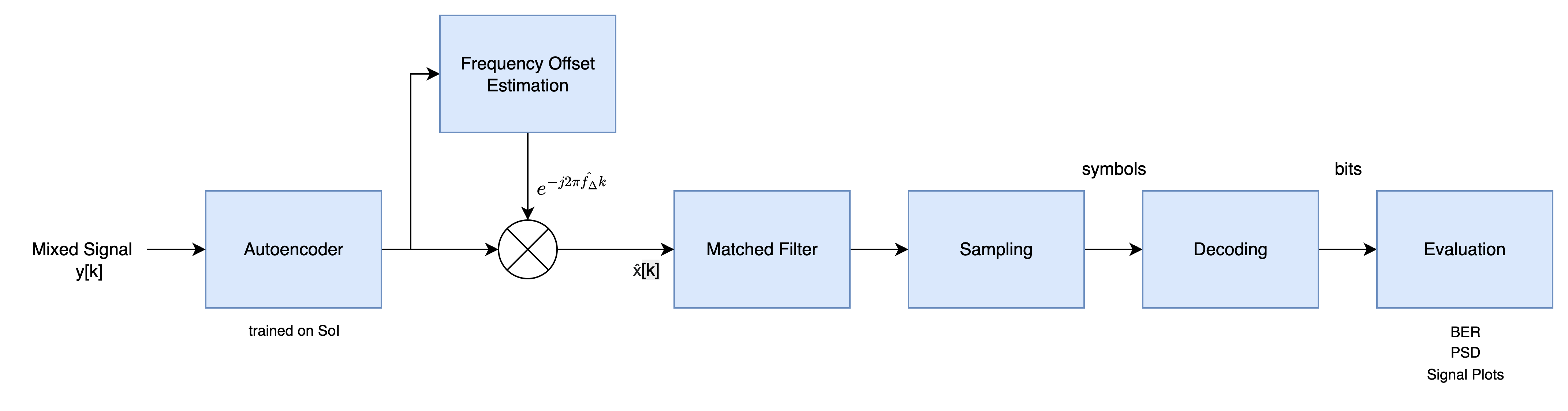}
    \caption{QPSK with Frequency Offset and AWGN Receiver Pipeline}
    \label{fig:qpsk_cfo_system}
\end{figure}

The U-Net model in this scenario is trained to denoise the AWGN and estimate the offsetted signal. A frequency estimation algorithm is then used on the output of the autoencoder. \Cref{fig:qpsk_cfo_system} shows the system used to obtain the bits in this scenario.

For frequency offset estimation, the $M^{\text{th}}$-power algorithm is applied. This technique exploits the rotational symmetry of the QPSK constellation by raising the received signal to the $M^{\text{th}}$ power, where $M$ is the constellation order (in this case, $M=4$). The frequency offset is then estimated from the phase of the resulting sequence and compensated prior to detection.

\begin{figure}[!htb]
    \centering
    \includegraphics[scale=0.5]{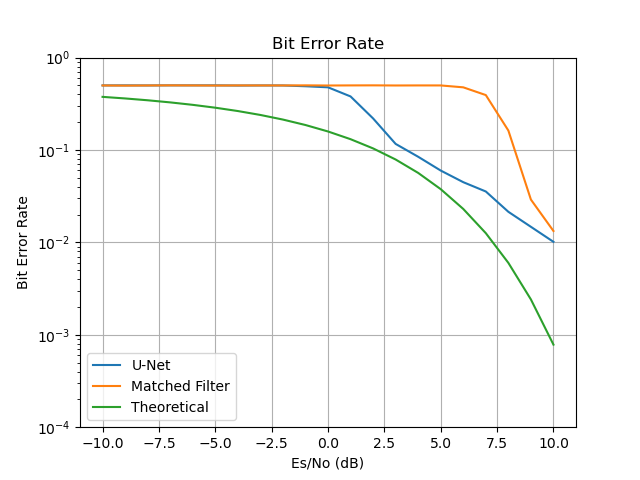}
    \caption{BER across Es/No of QPSK with Frequency Offset + AWGN} 
    \label{fig:qpsk_cfo_ber}
\end{figure}

\begin{figure}[!htb]
    \centering
    \includegraphics[scale=0.5]{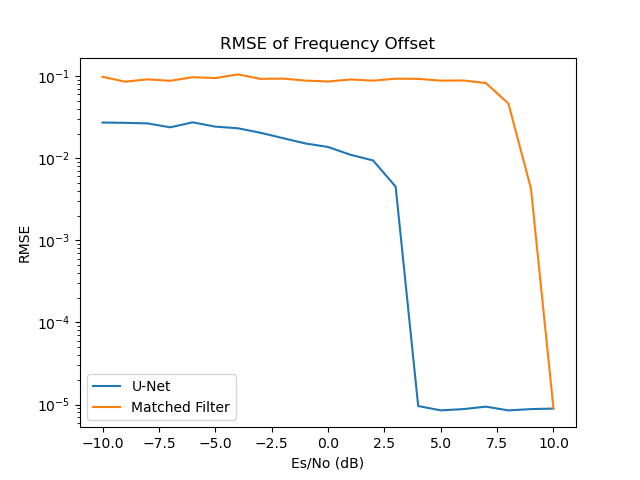}
    \caption{RMSE of Frequency Offset Estimation of QPSK with Frequency Offset + AWGN} 
    \label{fig:qpsk_cfo_fo}
\end{figure}

Matched filtering, which assumes perfect synchronization, is unable to fully correct for the distortion caused by CFO. As a result, constellation points drift into the wrong decision region, increasing the bit error rate (BER). In contrast, the proposed autoencoder learns to implicitly compensate for the noise leading to better frequency offset estimation during reconstruction. This can be evidently seen in BER and Frequency Offset Error plots in \Cref{fig:qpsk_cfo_ber,fig:qpsk_cfo_fo}.

\begin{figure}[!htb]
    \centering
    \includegraphics[scale=0.5]{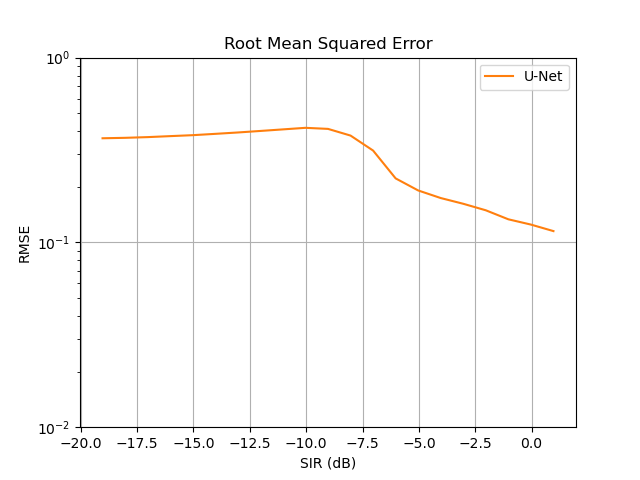}
    \caption{Root Mean Square Error across Es/No for QPSK with Frequency Offset + AWGN} 
    \label{fig:mse_qpsk_cfo}
\end{figure}

The BER plots confirm that the autoencoder outperforms matched filtering beyond 0dB Es/No. At Es/No less than 0dB, the autoencoder is not able to estimate the single frequency offset present in the mixed signal because of the stronger wideband noise component, but beyond 0dB it outperforms the matched filter and quickly reaches 0 offset error at 3dB Es/No. This demonstrates the strength of data-driven approaches in environments where classical assumptions such as perfect synchronization do not hold. \Cref{fig:qpsk_cfo_fo} shows the error in the frequency estimation on the the noisy signals filtered through matched filter and the autoencoder. Because the model is trained on random frequency offsets there is a residual phase shift introduced in the symbols. \Cref{fig:mse_qpsk_cfo} shows the Root Mean Squared Error does not improve till 0dB post which the U-Net module is able to denoise well.

% \begin{figure}[!htb]
% \centering
% \begin{subfigure}{.5\textwidth}
%   \centering
%   \includegraphics[scale=0.5]{figures/qpsk_cfo/new/ber.png}
%   \caption{Bit Error Rate across Es/No} 
%   \label{sfig:qpsk_cfo_ber}
% \end{subfigure}%
% \begin{subfigure}{.5\textwidth}
%   \centering
%   \includegraphics[scale=0.5]{fig/qpsk_cfo/new/freq_offset.png}
%   \caption{RMSE of Frequency Offset Estimation} 
%   \label{sfig:qpsk_cfo_fo}
% \end{subfigure}
% \caption{Evaluation of QPSK with Frequency Offset + AWGN}
% \label{fig:qpsk_cfo_performance}
% \end{figure}

\subsection{QPSK and Colored Noise}

In this section, the autoencoder-based mitigation framework is evaluated in the presence of colored noise impairments. Unlike AWGN, which is spectrally flat and uncorrelated, colored noise introduces temporal correlation or high-energy bursts that degrade the performance of classical receivers. Two representative types of colored noise are considered: AR(1) noise and $1/f$ noise.

\begin{figure}[htb]
    \centering
    \includegraphics[scale=0.5]{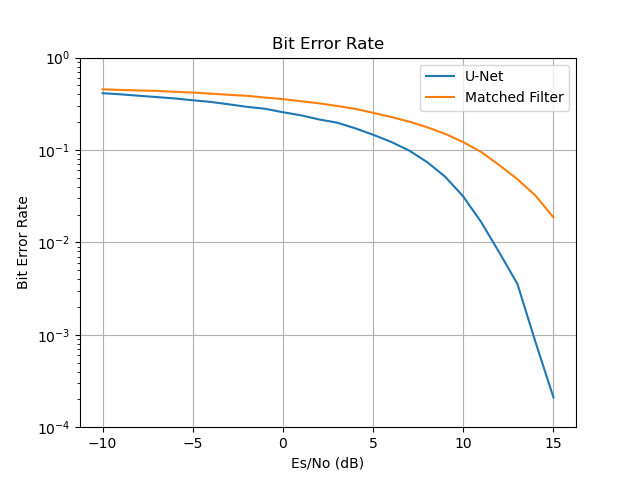 }
    \caption{Bit Error Rate across Es/No for QPSK + AR1 Noise} 
    \label{fig:ber_qpsk_color1}
\end{figure}

\paragraph{AR(1) Noise}: 
The presence of AR(1) noise creates temporal correlation across samples, distorting the effective noise distribution seen by the receiver. Matched filtering, which is designed for uncorrelated Gaussian noise, shows degradation in bit error rate (BER) compared to the AWGN case. In contrast, the autoencoder demonstrates the ability to learn and partially whiten the temporally correlated noise, resulting in lower BER across the evaluated $E_s/N_0$ range [\Cref{fig:ber_qpsk_color1}].

\begin{figure}[!htb]
    \centering
    \includegraphics[scale=0.5]{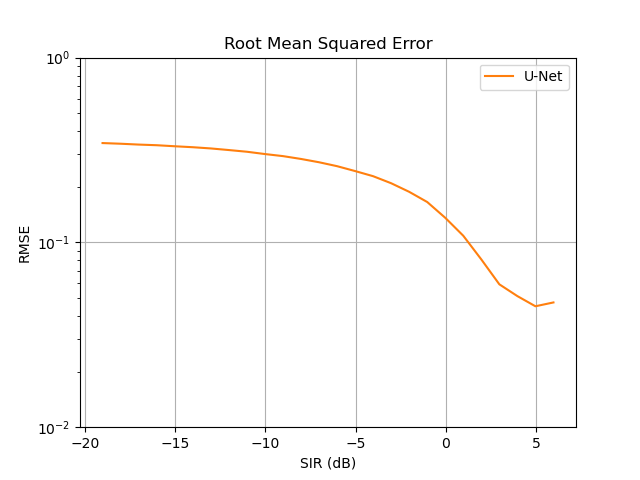}
    \caption{Root Mean Square Error across Es/No for QPSK + AR1 Noise} 
    \label{fig:mse_qpsk_color1}
\end{figure}

\begin{figure}[htb]
    \centering
    \includegraphics[scale=0.5]{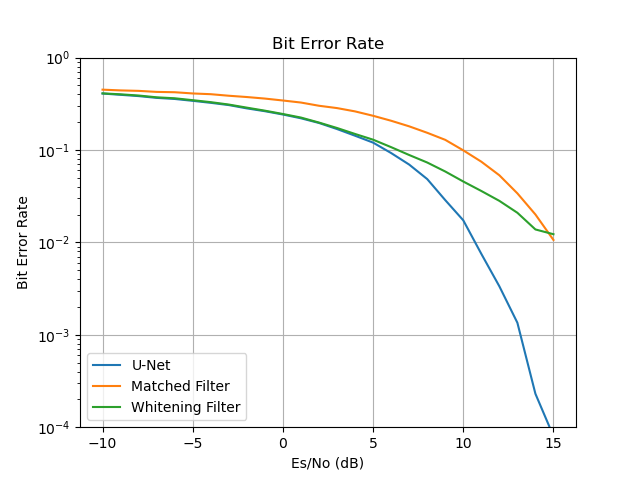}
    \caption{Bit Error Rate across Es/No for QPSK + 1/f Noise} 
    \label{fig:ber_qpsk_color2}
\end{figure}

\paragraph{$1/f$ Noise}:
The $1/f$ noise, which concentrates energy at low frequencies, significantly impacts the baseband spectrum of the QPSK signal. Constellation diagrams after matched filtering reveal substantial distortion in symbol placement, especially for low $E_s/N_0$. The matched filter is no longer the optimal filter for this colored noise so a whitening filter was applied and the whitened output was matched filtered to obtain the bits. The autoencoder is able to attenuate the low-frequency excess energy and reconstruct constellation points more closely aligned with the true signal. BER plots confirm improved detection accuracy compared to the matched filter as well as the whitened filter baseline [\Cref{fig:ber_qpsk_color2}].

\begin{figure}[!htb]
    \centering
    \includegraphics[scale=0.5]{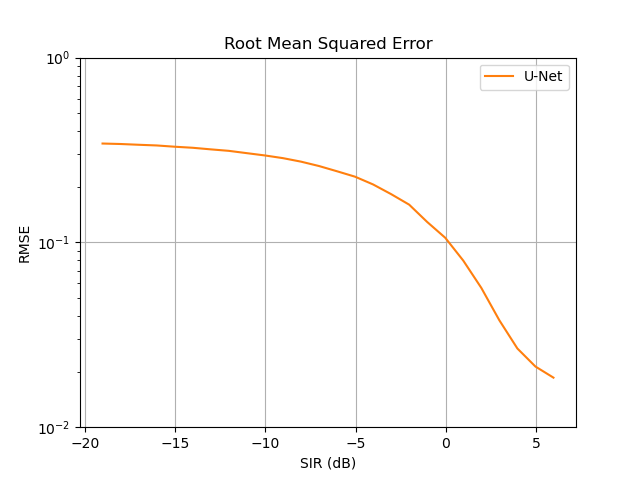}
    \caption{Root Mean Square Error across Es/No for QPSK + 1/f Noise} 
    \label{fig:mse_qpsk_color2}
\end{figure}

\subsection{QPSK and Impulsive Noise}
\begin{figure}[!htb]
    \centering
    \includegraphics[scale=0.5]{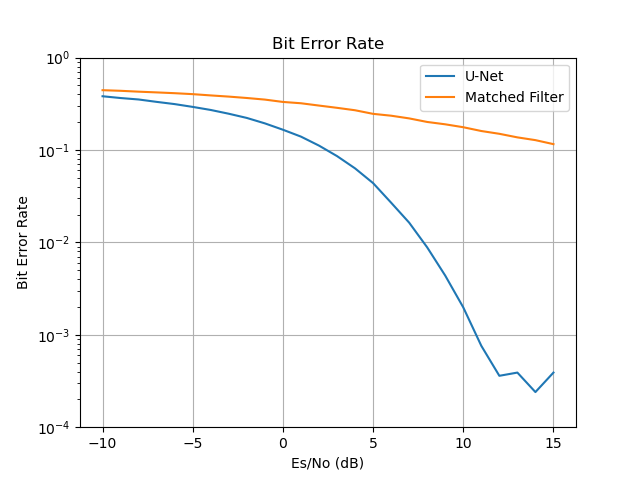}
    \caption{Bit Error Rate across Es/No for QPSK + Impulsive Noise} 
    \label{fig:ber_qpsk_color3}
\end{figure}

Impulsive noise produces occasional large-amplitude outliers, which can cause burst errors when passed through conventional receivers. In the time domain, this is observed as spikes in the received waveform, while in the frequency domain, impulsive noise contributes to broadband distortion. Matched filtering is highly sensitive to these impulses, resulting in error floors at high $E_s/N_0$. The autoencoder, trained on examples with impulsive events, learns to suppress these bursts while preserving the QPSK waveform structure. As a result, the autoencoder achieves significantly lower BER and cleaner constellation recovery compared to the matched filter which can be seen in \Cref{fig:ber_qpsk_color3}.  

\begin{figure}[!htb]
    \centering
    \includegraphics[scale=0.5]{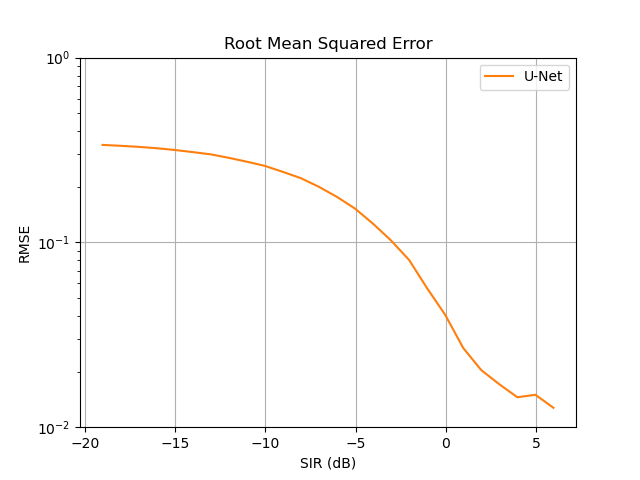}
    \caption{Root Mean Square Error across Es/No for QPSK + Impulsive Noise} 
    \label{fig:mse_qpsk_color3}
\end{figure}

Across all three non-AWGN scenarios, the autoencoder consistently outperforms the matched filter baseline.  The gains of the autoencoder are most pronounced for impulsive noise, where the non-Gaussian characteristics lead to severe degradation of conventional methods. These results demonstrate the adaptability of deep learning approaches in handling complex, non-ideal noise environments that deviate from the AWGN assumption. \Cref{fig:mse_qpsk_color1,fig:mse_qpsk_color2,fig:mse_qpsk_color3} shows the RMSE for the non-AWGN scenarios.

The matched filter is optimal only for AWGN. For non-white or impulsive noise, it no longer maximizes SNR because the noise is correlated or heavy-tailed. Impulsive noise is typically not white since its occurrence is sporadic and its energy is unevenly distributed across time, leading to temporal correlation and non-uniform spectral density. A whitening filter was developed for AR1 noise but the performance showed worse bit error rate owing to incorrect estimation of rho. For impulsive noise, a median filter was implemented to reduce the effects of burst noise but it did not outperform the matched filtered bit error rate.

\subsection{QAM and AWGN}

\begin{figure}[!htb]
    \centering
    \includegraphics[scale=0.5]{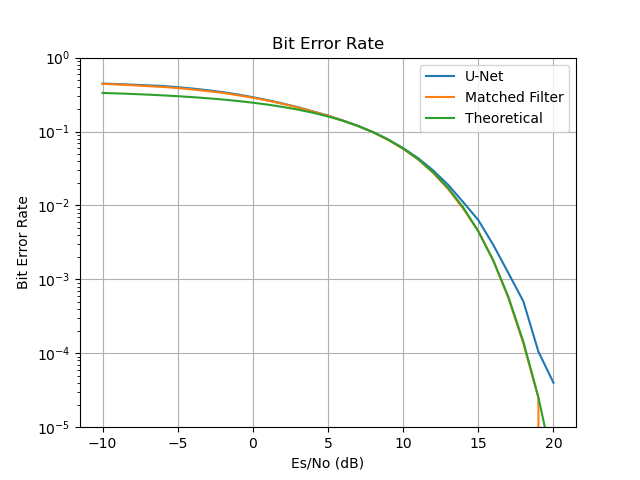}
    \caption{Bit Error Rate across Es/No for QAM + AWGN} 
    \label{fig:ber_qam}
\end{figure}

\begin{figure}[!htb]
    \centering
    \includegraphics[scale=0.5]{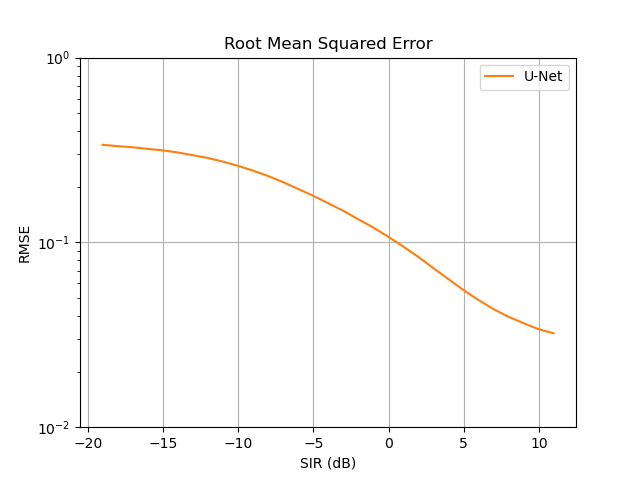}
    \caption{Root Mean Square Error across Es/No for QAM + AWGN} 
    \label{fig:mse_qam}
\end{figure}

In addition to QPSK-based interference mitigation, this work also explores the case of Quadrature Amplitude Modulation (QAM) signals corrupted by Additive White Gaussian Noise (AWGN). The QAM + AWGN scenario represents a more complex signal environment compared to constant-envelope modulations like QPSK, as it introduces both amplitude and phase variations that are more susceptible to noise distortion. Studying this case provides insight into the robustness and generalization ability of the proposed deep learning–based interference mitigation framework when applied to higher-order modulation schemes. As observed in the QPSK case, the model is able to follow the optimal matched filter performance for the QAM modulated signal as well which can be observed in \Cref{fig:ber_qam,fig:mse_qam}.

\section{Signal, Interference, and Noise}\label{sec:s_i_n}

\subsection{QPSK SoI with Sinusoid Interference and AWGN}

In this experiment, the signal of interest is a QPSK waveform corrupted by a narrowband sinusoidal interferer and additive white Gaussian noise. Two subcases are considered: (i) the sinusoidal tone lies within the passband of the QPSK signal (\textit{in-band}), and (ii) the sinusoid lies outside the passband (\textit{out-of-band}). The strength of the interference is controlled by the Signal-to-Interference Ratio (SIR), which is varied from $-10$~dB to $+10$~dB, while the noise level is fixed at $E_s/N_0 = 10$~dB. 

\subsubsection{In-band Interference.}  
\begin{figure}[!htb]
    \centering
    \includegraphics[scale=0.5]{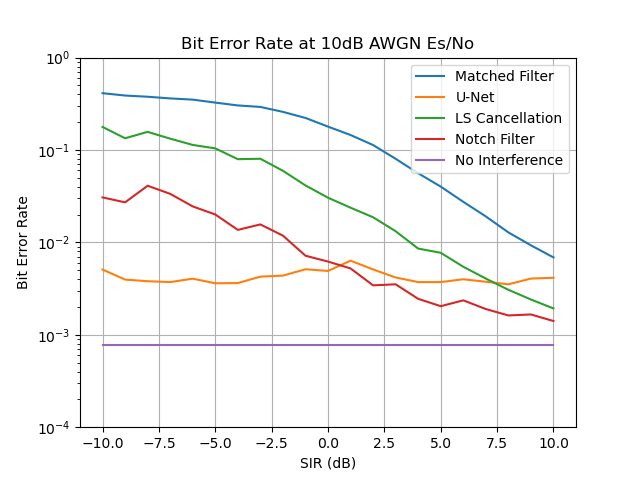}
    \caption{Bit Error Rate across SIR for QPSK + Single Tone In-Band Interference} 
    \label{fig:ber_qpsk_singletone1}
\end{figure}

When the sinusoidal interferer falls within the spectral support of the QPSK signal, it produces strong corruption that overlaps with the useful information-bearing frequencies. Conventional matched filtering struggles under this condition, as the narrowband interference passes through the filter without significant attenuation until the QPSK is much stronger than the Interference. The autoencoder, on the other hand, learns to suppress the strong sinusoidal component in the time domain while retaining the QPSK structure. BER results confirm a clear performance gain compared to the matched filter, particularly when the interference is dominant. To benchmark the proposed autoencoder-based approach, two conventional narrowband cancellation techniques: Least Squares (LS) estimation and the Notch filter, were also implemented for comparison. The Notch filter demonstrates slightly better performance than the autoencoder at positive SIR values, effectively suppressing the interference when it is relatively strong. However, the autoencoder-based method consistently outperforms LS estimation-based cancellation up to around 7 dB SIR, highlighting its robustness in low and moderate interference regimes where adaptive linear methods struggle to generalize. \Cref{fig:ber_qpsk_singletone1} shows the performance of the filters and autoencoder in presence of the single-tone interference as well as the expected BER in the absence of interference. The autoencoder is able to suppress the single-tone peak completely for both of the extreme cases and is able to reduce the overall noise-spread with few outlier symbols caused by the limitation of the autoencoder.

\subsubsection{Out-of-band Interference}  
\begin{figure}[!htb]
    \centering
    \includegraphics[scale=0.5]{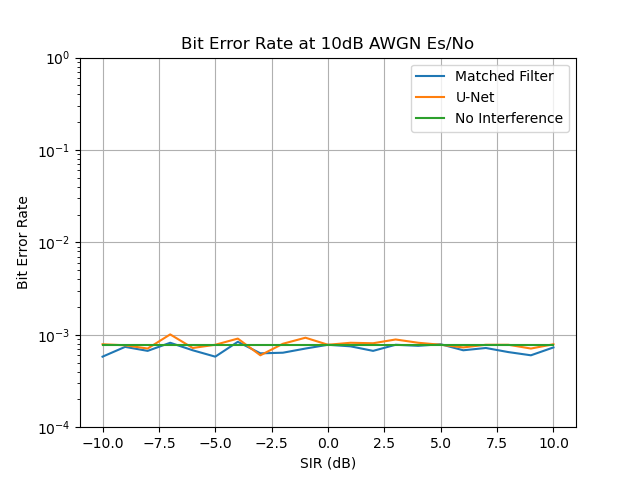}
    \caption{Bit Error Rate across SIR for QPSK + Single Tone Out-of-Band Interference} 
    \label{fig:ber_qpsk_singletone2}
\end{figure}

When the sinusoidal tone lies outside the QPSK passband, both matched filtering and the U-Net autoencoder provide effective rejection of the interference. The frequency selectivity of the RRC matched filter attenuates the out-of-band sinusoid significantly, leaving only minor residual leakage. Since the U-Net is trained to reproduce the clean QPSK signal, its output closely matches that of the matched filter in this case. BER performance for both approaches follows the theoretical bound for QPSK in AWGN [\Cref{fig:ber_qpsk_singletone2}], and constellation recovery is nearly ideal across the full SIR range and the noise variance is much lower than the matched filter approach. The autoencoder is successful at creating a high-pass filter for the out-of-band tone. As a result, there is no significant distinction between the two methods for the out-of-band case.

\subsection{QPSK SoI with LFM Interference and AWGN}

In this experiment, the signal of interest is a QPSK waveform corrupted by a wideband Linear Frequency Modulated (LFM) chirp interferer and additive white Gaussian noise (AWGN). The interference is scaled to achieve SIR values ranging from $-10$dB to $+10$dB, while the SoI is transmitted at $E_s/N_0 = 10$~dB. The LFM interferer introduces time-varying spectral overlap with the QPSK signal, producing a highly challenging interference scenario for conventional receivers.

\begin{figure}[!htb]
    \centering
    \includegraphics[scale=0.5]{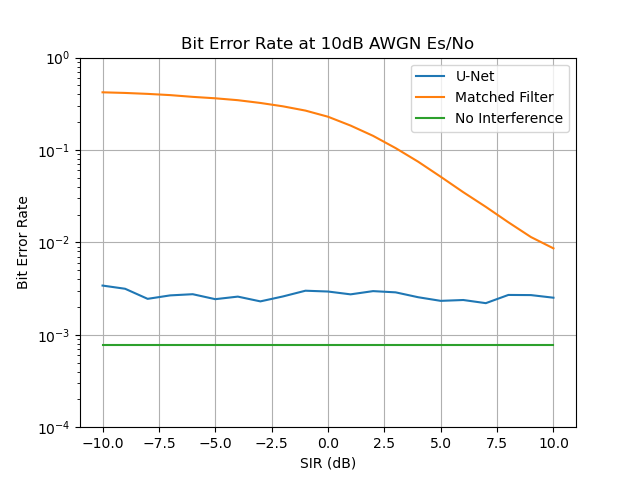}
    \caption{Bit Error Rate across SIR for QPSK + LFM Interference} 
    \label{fig:ber_qpsk_lfm}
\end{figure}

Classical matched filtering is weak against wideband LFM interference. Because the chirp sweeps through the entire baseband spectrum, the matched filter output contains significant residual interference energy, which manifests as distorted constellation points and elevated bit error rates. BER versus SIR curve in \Cref{fig:ber_qpsk_lfm} show that the matched filter performance degrades rapidly as the interference power increases, with detection nearly impossible at negative SIR values. The U-Net autoencoder demonstrates strong robustness in the presence of the LFM chirp. By operating directly on the I/Q time series, the autoencoder learns to suppress the structured, time-varying interference while preserving the symbol transitions of the QPSK signal. Time-domain reconstructions show effective removal of the chirp component, and power spectral density (PSD) comparisons confirm significant attenuation of the wideband interference energy. Constellation diagrams from the autoencoder output recover well-formed symbol clusters even at moderate SIR levels where the matched filter collapses.

\subsection{Successive Interference Cancellation (SIC)}
  
The successive interference cancellation method is employed in scenarios where the interference power dominates the signal of interest (SoI), that is, at low Signal-to-Interference Ratios. In this approach, the mixed signal is first passed through a matched filter designed for the interference waveform, typically an RRC filter matched to the interfering QPSK pulse shape. The filter output is then sampled to recover the interference symbols, which are demodulated into bits and subsequently remodulated to reconstruct the interference signal. To ensure proper cancellation, the reconstructed interference is scaled according to the estimated SIR before being subtracted from the original mixed signal. The resulting residual signal ideally contains only the SoI and noise. This residual is then processed by a matched filter corresponding to the SoI’s pulse shaping filter, sampled at the symbol instants, and demodulated to recover the transmitted SoI bits. 

Conversely, in scenarios where the SoI was dominant (SoI stronger than interference), the mixed signal was directly matched-filtered using the SoI’s RRC filter and demodulated without performing interference subtraction since we are only interested in the bit recovery of SoI.

\begin{figure}[!htb]
    \centering
    \includegraphics[width=1\linewidth]{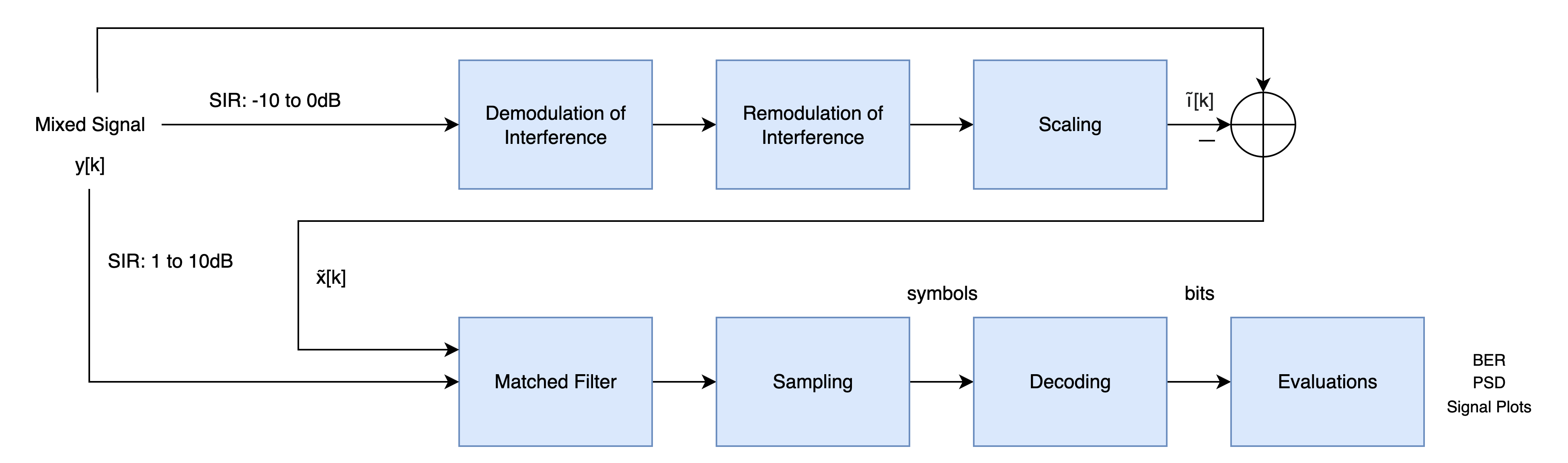}
    \caption{System Model for Successive Interference Cancellation}
    \label{fig:sic_system}
\end{figure}

While effective under moderate conditions, the performance of SIC relies heavily on accurate estimation of the interference symbols and the SIR; errors in this stage can propagate and reduce the accuracy of the SoI detection. \Cref{fig:sic_system} shows the pipeline used to implement SIC in this study. This marks as the second baseline along with the \Cref{fig:single_mf_system} for evaluating performance of the autoencoder approach.

\subsection{System Model}

For the cases where both the signal of interest and the interference have similar modulation scheme and structure (QPSK and QPSK), a novel successive U-Net system was develop which was devised as following: 

We consider the discrete-time complex baseband mixture
\[
y[k] = x[k] + i[k] + n[k],
\]
where \(x[k]\) is the pulse-shaped QPSK signal of interest (SoI), \(i[k]\) is the pulse-shaped QPSK interference generated independently of \(x[k]\), and \(n[k] \sim \mathcal{CN}(0,\sigma^2)\) is complex Gaussian noise. The samples per symbol are denoted \(sps\) and the matched filter at the receiver is the RRC filter \(h_{\text{RRC}}[k]\) that matches the transmit shaping.

The receiver uses two stages of autoencoders in sequence, together with a demodulate and remodulate step for power controlled subtraction. The first autoencoder is trained to estimate the interference from the noisy mixture. The second autoencoder is trained to estimate the SoI from a partially cleaned residual.

\subsubsection{Stage 1: Interference estimation}
The first autoencoder takes the mixture as input and outputs an interference estimate
\[
\hat{i}_{\text{ae}}[k] = \mathcal{A}_{\text{int}}\{y[k]\}.
\]
During training, \(\mathcal{A}_{\text{int}}\) is supervised with clean interference targets \(i[k]\) using a waveform MSE loss.

\subsubsection{Demodulate and remodulate, power normalization, and subtraction}
To reduce bias due to residual noise and encoder artifacts, \(\hat{i}_{\text{ae}}[k]\) is passed through a decision directed loop that demodulates to hard symbols and then remodulates with the known pulse shape. Let \(\mathcal{D}(\cdot)\) denote symbol decisions after matched filtering and timing recovery, and let \(\mathcal{M}(\cdot)\) denote remodulation with the known QPSK mapper and transmit RRC filter. The interference template is
\[
\tilde{i}[k] = \mathcal{M}\big(\mathcal{D}(\hat{i}_{\text{ae}}[k])\big).
\]

In this case, we assume that perfect knowledge about the SIR is known to us. We scaled the remodulated signal with the ground truth SIR with a factor of $\hat\alpha$
The first stage residual after cancellation is
\[
\tilde{x}[k] = y[k] - \hat{\alpha}\,\tilde{i}[k].
\]

\subsubsection{Stage 2: SoI estimation}
The second autoencoder operates on the residual and outputs a cleaned SoI estimate
\[
\hat{x}[k] = \mathcal{A}_{\text{soi}}\{\tilde{x}[k]\}.
\]
During training, \(\mathcal{A}_{\text{soi}}\) is supervised with clean SoI targets \(x[k]\) using a waveform MSE loss.

\subsubsection{Matched filtering, symbol decisions, and BER}
The estimate \(\hat{x}[k]\) is matched filtered and sampled at the symbol times
\[
z[m] = \big(\hat{x} * h_{\text{RRC}}\big)[m\,sps],
\]
then mapped to symbol decisions \(\hat{s}[m] = \mathcal{D}(z[m])\) and to bits \(\hat{b}\). The bit error rate is computed against the ground truth bits \(b\)
\[
\text{BER} = \frac{1}{|b|}\sum_{u} \mathbbm{1}\{\hat{b}_u \ne b_u\}
\]

The two autoencoders are trained with paired data. For \(\mathcal{A}_{\text{int}}\) the inputs are mixtures \(y[k]\) and targets are clean \(i[k]\). For \(\mathcal{A}_{\text{soi}}\) the inputs are residuals $\tilde{x[k]}$ formed by subtracting the mixed signal \(y[k]\) and scaled-remodulated $\tilde{i}[k]$, and targets are clean \(x[k]\). If carrier or timing offsets are present, the demodulate and remodulate block will include standard correction steps, for example \(M\)th power frequency estimation applied to \(\hat{i}_{\text{ae}}\) as well as to $\hat{x}$ as per the situation before decisions in a similar way that was done in \Cref{fig:qpsk_cfo_system}. This model is hereby referred to as \textbf{Successive Interference Cancellation using U-Net} or \textbf{SICU-Net}.

\begin{figure}[!htb]
    \centering
    \includegraphics[width=1\linewidth]{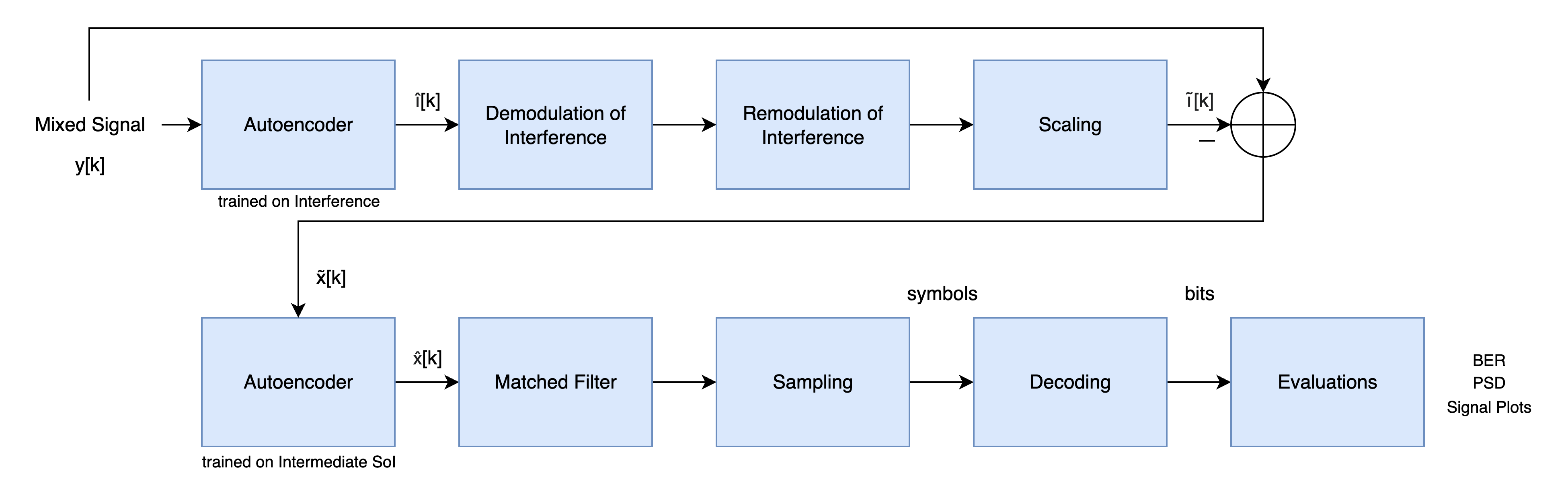}
    \caption{System Model for Successive Interference Cancellation using U-Net}
    \label{fig:sicunet}
\end{figure}

\subsection{QPSK SoI, QPSK Interference, and AWGN }\label{subsec:qpsk_qpsk}

This case represents one of the most challenging interference scenarios because both signals share the same modulation format and have overlapping spectral content. Unlike sinusoidal or LFM interference, which possess distinct spectral signatures, QPSK interference appears nearly identical to the desired    in the frequency domain. The ability to distinguish between them depends heavily on differences in bandwidth, timing, or power level, quantified by the Signal-to-Interference Ratio (SIR).  

Classical approaches such as matched filtering fail to reliably separate the SoI from QPSK interference, as both signals project onto similar basis functions. Successive interference cancellation (SIC) methods can mitigate the problem if the interference can be estimated and subtracted, but their performance is highly dependent on accurate demodulation of the interfering signal, which is difficult at low SIR.  

By contrast, the proposed autoencoder operates directly on raw I/Q sequences and learns to exploit subtle temporal and spectral differences between the SoI and the interfering QPSK. Even in conditions of strong interference or low SIR, the autoencoder is able to reconstruct a cleaner estimate of the SoI. Performance evaluation is carried out using bit error rate (BER) versus SIR curves, constellation recovery, and spectral analyses comparing the predicted signals to both the matched filter output and the ground-truth waveforms.

\begin{figure}[!htb]
    \centering
    \includegraphics[scale = 0.5]{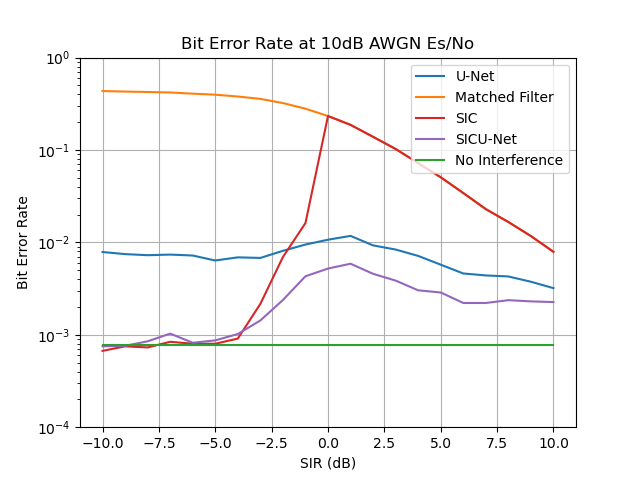}
    \caption{Bit Error Rate for QPSK + QPSK Interference (SPS 32)}
    \label{fig:ber_qpsk32}
\end{figure}

The samples per symbol (SPS) for the signal of interest's upsampling as well as filter has been fixed at 8 throughout the following case studies. The interference SPS is varied in-order to study the performance impact of different relative bandwidths. The bandwidth changes by a factor of 2 with a change of factor of 2 in SPS. Three Interference SPS: 32, 16, and 4 are tested in this case along with the fixed SoI SPS of 8. \Cref{fig:ber_qpsk32} shows the performance of both the systems  U-Net [\ref{fig:unet}] and SICU-Net [\ref{fig:sicunet}]. It can be observed that the successive cancellation approach is much more effective in suppressing the interference even in lower SIR scenarios where the interference heavily overpower the signal of interest. SICU-Net also works on-par with the traditional SIC approach in the lower SIR's and then displays a drastic improvement for SIR greater than -4dB. The performance of the autoencoder as well as traditional SIC degrades as the bandwidth of SoI gets closer to the bandwidth of the Interference.

\begin{figure}[!htb]
    \centering
    \includegraphics[scale = 0.5]{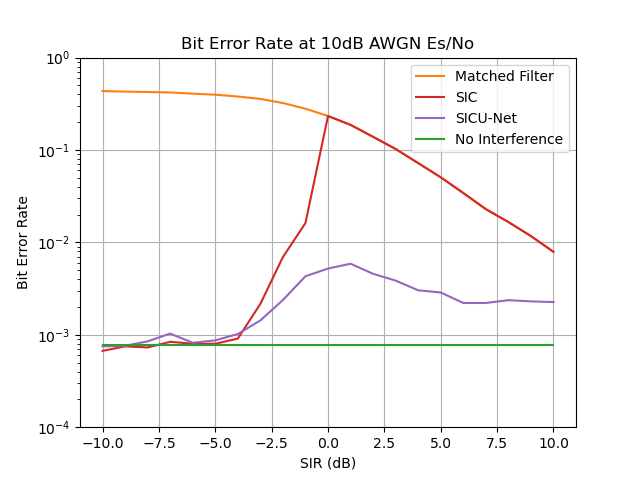}
    \caption{Bit Error Rate for QPSK + QPSK Interference (SPS 32)}
    \label{fig:ber_qpsk1}
\end{figure}
\begin{figure}[!htb]
    \centering
    \includegraphics[scale = 0.5]{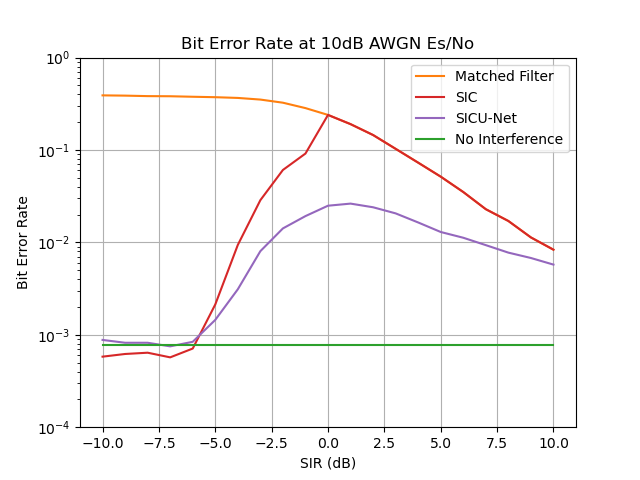}
    \caption{Bit Error Rate for QPSK + QPSK Interference (SPS 16)}
    \label{fig:ber_qpsk2}
\end{figure}
\begin{figure}[!htb]
    \centering
    \includegraphics[scale = 0.5]{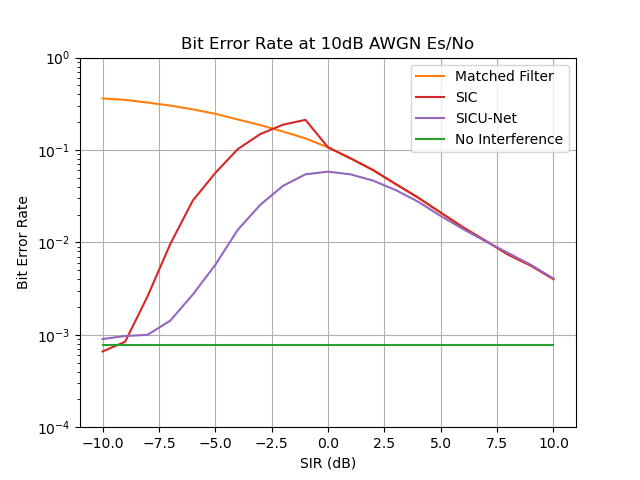}
    \caption{Bit Error Rate for QPSK + QPSK Interference (SPS 4)}
    \label{fig:ber_qpsk3}
\end{figure}

\subsection{QPSK SoI, QPSK Interference with Frequency Offset, and AWGN} \label{subsec:qpsk_qpsk_fo}

This scenario investigates the performance of the proposed two-stage autoencoder successive interference cancellation system in the presence of QPSK interference that is corrupted by carrier frequency offset (CFO). Four distinct subcases are considered to capture both fixed and random offset conditions.

\begin{figure}[!htb]
    \centering
    \includegraphics[width=1\linewidth]{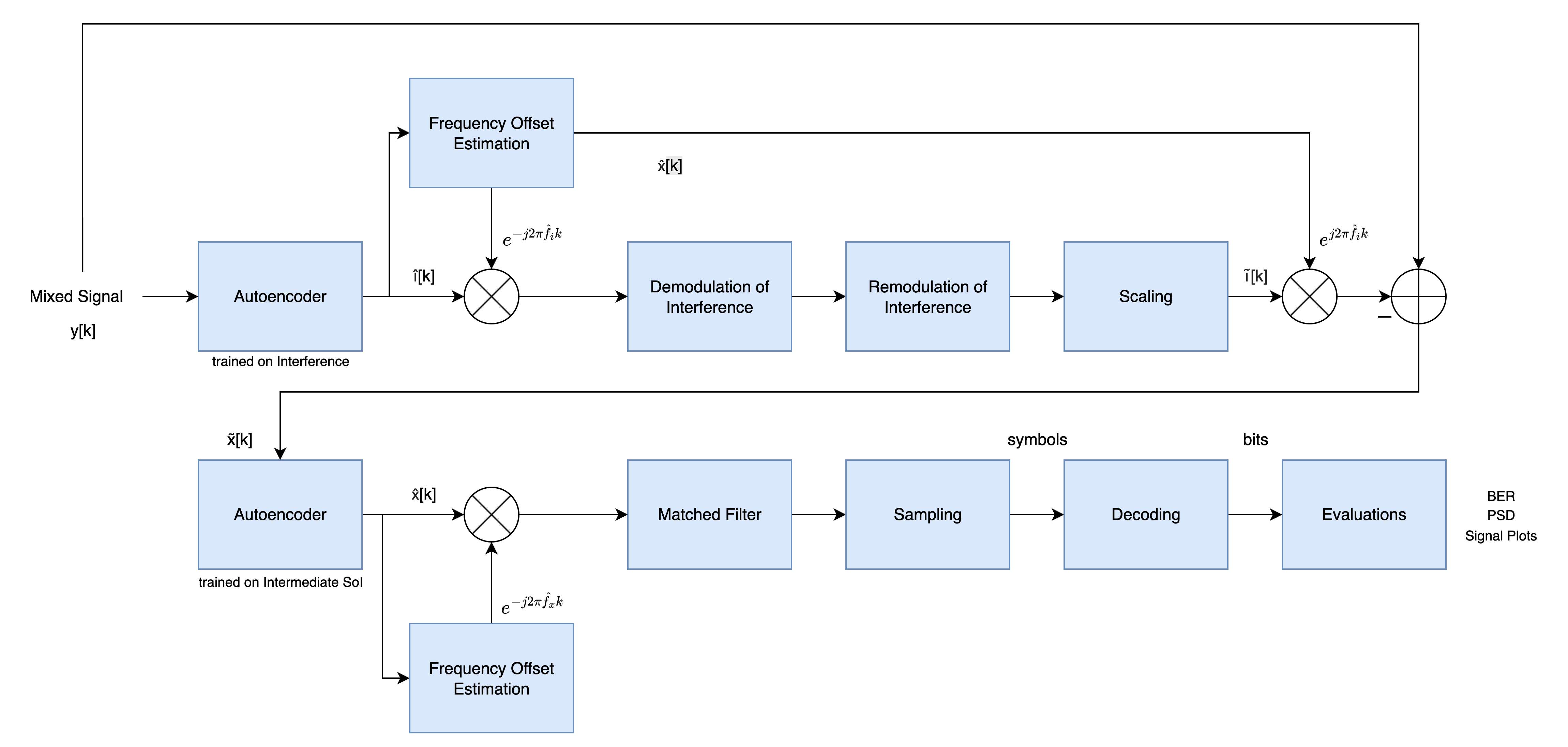}
    \caption{System Model for Successive Interference Cancellation using U-Net with Frequency Offset Estimation}
    \label{fig:sicunet_frequency}
\end{figure}

The received signal is modeled as
\[
y[k] = x[k] \, e^{j2\pi f_x k} + i[k] \, e^{j2\pi f_i k} + n[k],
\]
where $x[k]$ is the QPSK signal of interest (SoI), $i[k]$ is the interfering QPSK signal, $f_x$ and $f_i$ denote the normalized frequency offsets of the SoI and interference respectively, and $n[k] \sim \mathcal{CN}(0, \sigma^2)$ is additive Gaussian noise. \Cref{fig:sicunet_frequency} shows the receiver chain for these freqeuncy offset scenarios. The frequency offset estimator blocks are used depending on the presence of offset for the particular signal.

\begin{figure}[!htb]
    \centering
    \includegraphics[scale = 0.5]{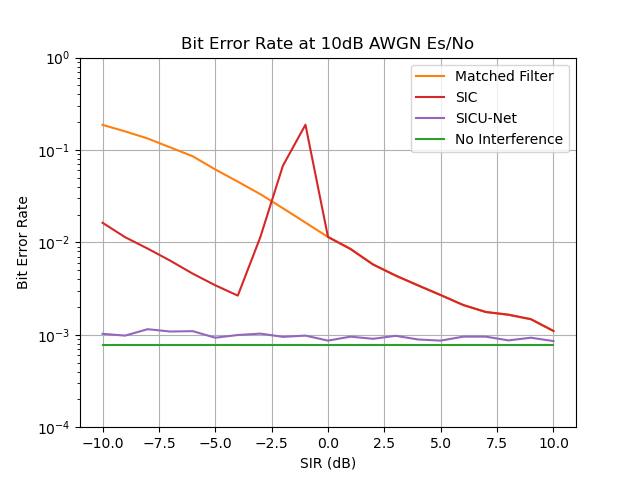}
    \caption{Evaluation of QPSK SoI where Interference includes a Frequency Offset (Fixed)}
    \label{fig:ber_qpsk_qpsk_cfo11}
\end{figure}

\paragraph{Interference-Only Frequency Offset (Fixed)}  
In the first subcase, only the interfering QPSK signal is subject to a frequency offset, while the SoI remains perfectly synchronized. The frequency offset $f_i$ is held constant across all realizations. \Cref{fig:ber_qpsk_qpsk_cfo11} shows the performance for this interference case. The normalized fixed frequency offset was kept as 0.08 cycles per sample.

% \begin{figure}[!htb]
%     \centering
%     \includegraphics[scale = 0.5]{figures/qpsk_qpsk_cfo/try1/fo_error.png}
%     \caption{Evaluation of QPSK SoI where Interference includes a Frequency Offset (Fixed)}
%     \label{fig:fo_qpsk_qpsk_cfo11}
% \end{figure}

% \begin{figure}[!htb]
% \centering
% \begin{subfigure}{.5\textwidth}
%   \centering
%   \includegraphics[scale=0.5]{fig/qpsk_qpsk_cfo/try1/new/ber.png}
%   \caption{Bit Error Rate} 
%   \label{fig:ber_qpsk_qpsk_cfo11}
% \end{subfigure}%
% \begin{subfigure}{.5\textwidth}
%   \centering
%   \includegraphics[scale=0.5]{fig/qpsk_qpsk_cfo/try1/new/fo_error.png}
%   \caption{RMSE of $f_i$} 
%   \label{fig:fo_qpsk_qpsk_cfo11}
% \end{subfigure}
% \caption{Evaluation of QPSK SoI where Interference includes a Frequency Offset (Fixed)}
% \label{fig:qpsk_qpsk_cfo11}
% \end{figure}

\begin{figure}[!htb]
    \centering
    \includegraphics[scale = 0.5]{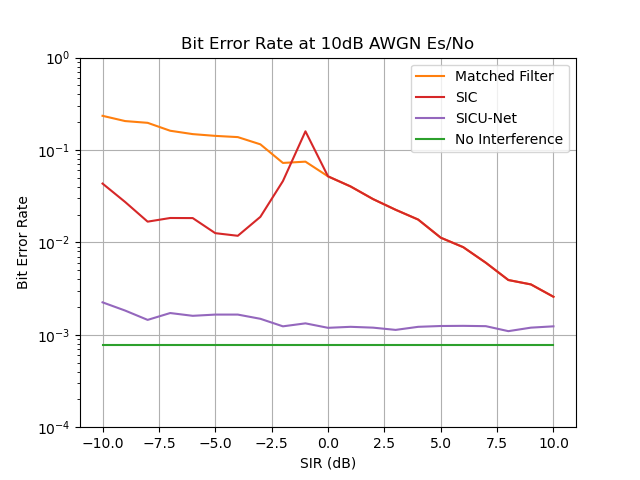}
    \caption{Evaluation of QPSK SoI where Interference includes a Frequency Offset (Random)}
    \label{fig:ber_qpsk_qpsk_cfo12}
\end{figure}

\paragraph{Interference-Only Frequency Offset (Random)}  
In the second subcase, the interference CFO $f_i$ is drawn randomly from a predefined range for each realization. This introduces variability in the interference spectrum and stresses the generalization ability of the autoencoder. \Cref{fig:ber_qpsk_qpsk_cfo12} shows the performance for this interference case. The normalized random frequency offset was picked from a uniform distribution between 0.05 - 0.1 cycles per sample.

% \begin{figure}[!htb]
% \centering
% \begin{subfigure}{.5\textwidth}
%   \centering
%   \includegraphics[scale=0.5]{fig/qpsk_qpsk_cfo/try2/new/ber.png}
%   \caption{Bit Error Rate} 
%   \label{fig:ber_qpsk_qpsk_cfo12}
% \end{subfigure}%
% \begin{subfigure}{.5\textwidth}
%   \centering
%   \includegraphics[scale=0.5]{fig/qpsk_qpsk_cfo/try2/new/fo_error.png}
%   \caption{RMSE of $f_i$} 
%   \label{fig:fo_qpsk_qpsk_cfo12}
% \end{subfigure}
% \caption{Evaluation of QPSK SoI where Interference includes a Frequency Offset (Random)}
% \label{fig:qpsk_qpsk_cfo12}
% \end{figure}

\begin{figure}[!htb]
    \centering
    \includegraphics[scale = 0.5]{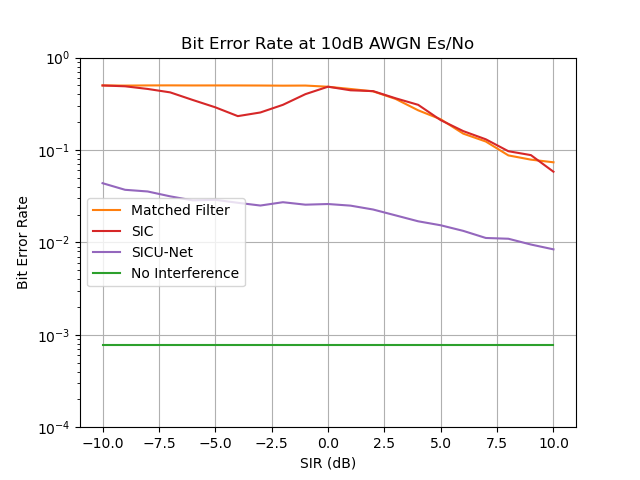}
    \caption{Performance for both SoI and Interference Frequency Offset (Fixed)}
    \label{fig:ber_qpsk_qpsk_cfo21}
\end{figure}

\paragraph{Both SoI and Interference Frequency Offset (Fixed)}  
In this subcase, both the SoI and the interference experience fixed frequency offsets, $f_x$ and $f_i$, applied consistently across realizations. The presence of CFO on the SoI complicates symbol detection even after interference suppression, as constellation points exhibit steady rotation. The demodulate–remodulate stage within the SIC framework partially compensates for this, but residual distortion remains. Results show that while the autoencoder pipeline continues to provide gains over matched filtering, BER curves are shifted upward compared to the interference-only CFO case. \Cref{fig:ber_qpsk_qpsk_cfo21} shows the performance for this interference case. The normalized frequency offset for interference was 0.08 cycles per sample and for SoI was 0.05 cycles per sample.

% \begin{figure}[!htb]
% \centering
% \begin{subfigure}{.3\textwidth}
%   \centering
%   \includegraphics[scale=0.3]{fig/qpsk_qpsk_cfo_2/try1/new/ber.png}
%   \caption{Bit Error Rate} 
%   \label{fig:ber_qpsk_qpsk_cfo21}
% \end{subfigure}%
% \begin{subfigure}{.3\textwidth}
%   \centering
%   \includegraphics[scale=0.3]{fig/qpsk_qpsk_cfo_2/try1/new/fo_error_i.png}
%   \caption{RMSE of $f_i$} 
%   \label{fig:foe_interference_21}
% \end{subfigure}%
% \begin{subfigure}{.3\textwidth}
%   \centering
%   \includegraphics[scale=0.3]{fig/qpsk_qpsk_cfo_2/try1/new/fo_error_s.png}
%   \caption{RMSE of $f_x$} 
%   \label{fig:foe_soi_21}
% \end{subfigure}
% \caption{Performance for Both SoI and Interference Frequency Offset (Fixed)}
% \label{fig:performance_qpsk_qpsk_cfo21}
% \end{figure}

\begin{figure}[!htb]
    \centering
    \includegraphics[scale = 0.5]{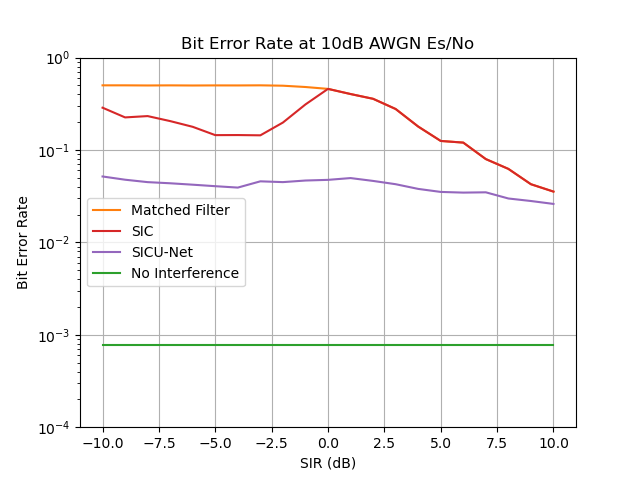}
    \caption{Performance for both SoI and Interference Frequency Offset (Random)}
    \label{fig:ber_qpsk_qpsk_cfo22}
\end{figure}

\paragraph{Both SoI and Interference Frequency Offset (Random)}  
The final subcase considers the most challenging condition, where both the SoI and interference are subject to randomly varying CFOs. Each realization has independent draws for $f_x$ and $f_i$ from the defined range. This produces nonstationary interference and signal distortions that are difficult to correct with classical methods. The autoencoder system demonstrates partial success in reconstructing the SoI, but BER performance shows increased variability across trials due to the compounded distortions. Nevertheless, the results remain consistently better than matched filter baselines, highlighting the adaptability of the proposed framework. \Cref{fig:ber_qpsk_qpsk_cfo22} shows the performance for this interference case. Random frequency offset was chosen from uniform distribution between 0.05 - 0.1 cycles per sample for both interference and SoI.

% \begin{figure}[!htb]
% \centering
% \begin{subfigure}{.3\textwidth}
%   \centering
%   \includegraphics[scale=0.3]{fig/qpsk_qpsk_cfo_2/try2/new/ber.png}
%   \caption{Bit Error Rate} 
%   \label{fig:ber_qpsk_qpsk_cfo22}
% \end{subfigure}%
% \begin{subfigure}{.3\textwidth}
%   \centering
%   \includegraphics[scale=0.3]{fig/qpsk_qpsk_cfo_2/try2/new/fo_error_i.png}
%   \caption{RMSE of $f_i$} 
%   \label{fig:foe_interference_22}
% \end{subfigure}%
% \begin{subfigure}{.3\textwidth}
%   \centering
%   \includegraphics[scale=0.3]{fig/qpsk_qpsk_cfo_2/try2/new/fo_error_s.png}
%   \caption{RMSE of $f_x$} 
%   \label{fig:foe_soi_22}
% \end{subfigure}
% \caption{Performance for both SoI and Interference Frequency Offset (Random)}
% \label{fig:performance_qpsk_qpsk_cfo22}
% \end{figure}

Across all four subcases, the proposed autoencoder-based SIC system provides improved performance compared to matched filtering. The frequency offset estimation of the autoencoder generated signal is close to perfect and error is zero. The system shows strong performance when only the interference is affected by CFO, even under random offset conditions. In this case, the SIC performance at SIR close to 0dB is worse than the direct MF approach because of incorrect frequency offset estimation which propagates through the system which can be seen as a hump. It is interesting to note that in the presence of frequency offset in the interference, the overall performance of the auto-encoder reaches near the theoretical no interference level which was not observed in the previous case where there was no frequency offset. When the interference signal has a frequency offset relative to the desired QPSK signal, the two signals become less correlated in frequency. This frequency separation makes the interference more distinguishable from the desired signal in both time and spectral domains. As a result, the model can more easily learn to isolate and reconstruct the interference component because its features differ more clearly from those of the target signal. In contrast, when there is no CFO, the interference and desired QPSK signals overlap perfectly in frequency and phase space. This makes their superposition highly structured and more difficult for both traditional and ML-based methods to separate, since their modulation patterns are similar and coherent. The introduction of the frequency offset allows for a better estimation and thereby better mitigation of the interference. When the SoI also suffers from CFO, performance degrades, but the autoencoder still provides meaningful gains over the baseline once the frequency offset estimation error increases. These findings emphasize the potential of data-driven interference mitigation to handle synchronization impairments that challenge conventional approaches.

\subsection{QPSK SoI, QPSK Interference with Timing Offset, and AWGN} \label{subsec:qpsk_qpsk_to}

In this scenario, the QPSK signal of interest (SoI) is corrupted by an interfering QPSK signal that is misaligned in time, together with additive white Gaussian noise (AWGN). The timing offset of the interference is modeled by prepending a sequence of zeros to the interferer prior to addition with the SoI. The received baseband signal can be expressed as
\[
y[k] = x[k] + \tilde{i}[k] + n[k],
\]
where $x[k]$ is the pulse-shaped QPSK SoI, $n[k] \sim \mathcal{CN}(0,\sigma^2)$ is complex Gaussian noise, and $\tilde{i}[k]$ is the interference with a timing offset $\tau$ samples introduced by
\[
\tilde{i}[k] =
\begin{cases}
0, & k < \tau, \\
i[k-\tau], & k \geq \tau,
\end{cases}
\]
with $i[k]$ denoting the original interference waveform.  

This misalignment creates symbol-level overlap between the SoI and the interfering signal that does not correspond to aligned constellation points. As a result, conventional successive interference cancellation (SIC), which relies on demodulating the interference accurately, suffers from residual errors after cancellation.

\begin{figure}[!htb]
    \centering
    \includegraphics[scale = 0.5]{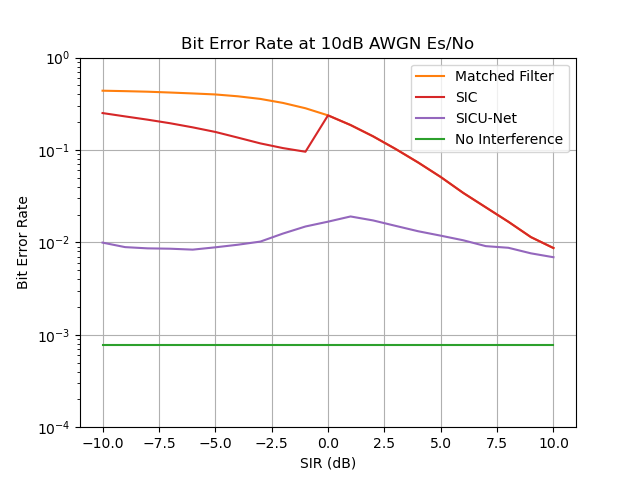}
    \caption{Bit Error Rate for Fixed $\tau$}
    \label{fig:ber_qpsk_qpsk_to1}
\end{figure}

\paragraph{Fixed Timing Offset}  
When the timing offset $\tau$ is fixed for all realizations, the autoencoder trained to predict the interference without the offset is able to reconstruct the non-offseted version of the interferer. After subtraction, the residual contains significantly less distortion compared to conventional SIC. BER versus SIR curves in \Cref{fig:ber_qpsk_qpsk_to1} confirm that the autoencoder-assisted approach consistently outperforms SIC in this case. $\tau$ was 40 samples for this scenario. 

\begin{figure}[!htb]
    \centering
    \includegraphics[scale = 0.5]{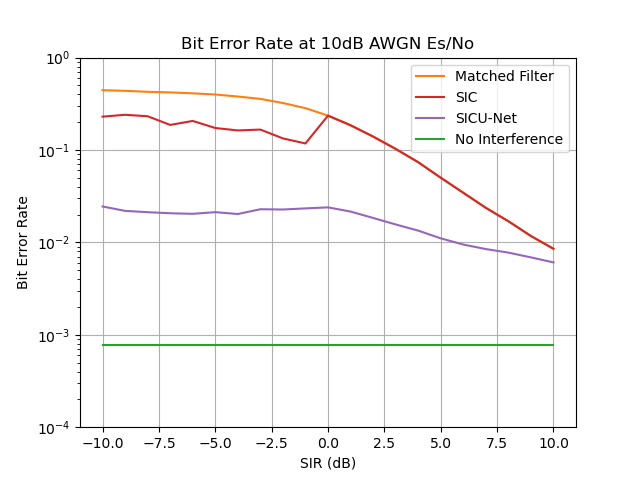}
    \caption{Bit Error Rate for Random $\tau$}
    \label{fig:ber_qpsk_qpsk_to2}
\end{figure}

\paragraph{Random Timing Offset}  
When the timing offset $\tau$ is drawn randomly from a specified range for each realization, the interferer is misaligned differently across examples. This randomness further degrades SIC performance, as the demodulate–remodulate stage cannot easily adapt to arbitrary shifts. The autoencoder, however, remains robust, producing an interference estimate that is approximately aligned to its canonical form regardless of the input padding. Although performance degrades slightly compared to the fixed-offset case, the BER results in \Cref{fig:ber_qpsk_qpsk_to2} still show a clear advantage over SIC. $\tau$ was randomly chosen from a uniform distribution of 0 to 50 samples. 

% \begin{figure}[!htb]
% \centering
% \begin{subfigure}{.5\textwidth}
%   \centering
%   \includegraphics[scale=0.5]{fig/qpsk_qpsk_to/try1/new/ber.png}
%   \caption{Bit Error Rate for Fixed $\tau$} 
%   \label{sfig:ber_qpsk_qpsk_to1}
% \end{subfigure}%
% \begin{subfigure}{.5\textwidth}
%   \centering
%   \includegraphics[scale=0.5]{fig/qpsk_qpsk_to/try2/new/ber.png}
%   \caption{Bit Error Rate for Random $\tau$} 
%   \label{sfig:ber_qpsk_qpsk_to2}
% \end{subfigure}
% \caption{Evaluation of QPSK SoI with Interference-Only Timing Offset}
% \label{fig:ber_qpsk_qpsk_to}
% \end{figure}

By modeling timing offset as zero-padding and training the autoencoder to output interference without the offset, the network effectively learns to invert the misalignment distortion. This allows the autoencoder-based system to achieve significantly better interference cancellation than SIC under both fixed and random timing offset conditions, demonstrating the adaptability of deep learning methods to synchronization-related impairments.

\subsection{QPSK SoI, QAM Interference, and AWGN}
\begin{figure}[!htb]
    \centering
    \includegraphics[scale = 0.5]{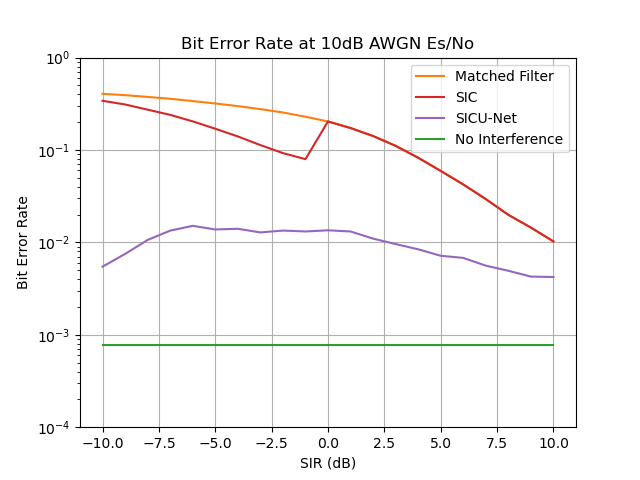}
    \caption{Bit Error Rate for QPSK + QAM Interference (SPS 32)}
    \label{fig:ber_qam32}
\end{figure}

Another scenario investigated in this work involves a mixed-signal environment consisting of a QPSK SoI and a QAM interference signal in the presence of AWGN. This QPSK + QAM + AWGN configuration is designed to evaluate the model’s ability to handle cross-modulation interference, where the signal of interest and the interfering signal follow different modulation formats. Such conditions are common in heterogeneous communication environments, where co-channel interference may arise from systems employing dissimilar modulation schemes. The introduction of QAM interference adds further complexity due to its amplitude variations and higher spectral efficiency, which can distort both the phase and amplitude characteristics of the QPSK SOI. This experiment provides valuable insights into the adaptability and discrimination capability of the proposed autoencoder-based interference mitigation framework when exposed to diverse modulation interference types. \Cref{fig:ber_qam32} shows that the bit error rate of QPSK SoI degrades with the presence of QAM interference when compared to the previous QPSK interference scenario. This translates to a worse interference estimation of the QAM symbols as it is an inherently harder task than QAM interference estimation. It can be observed that while at lower SIR where the QAM overpowers QPSK signal, the estimated symbols are in their well defined regions, but for higher SIR the interference is not being recreated accurately but since the SoI is stronger, the symbol demodulation is much better than the successive matched filtered method.

\section{Noise, Interference, and SIR Classifiers}\label{sec:nic}

\begin{figure}[!htb]
    \centering
    \includegraphics[width=1\linewidth]{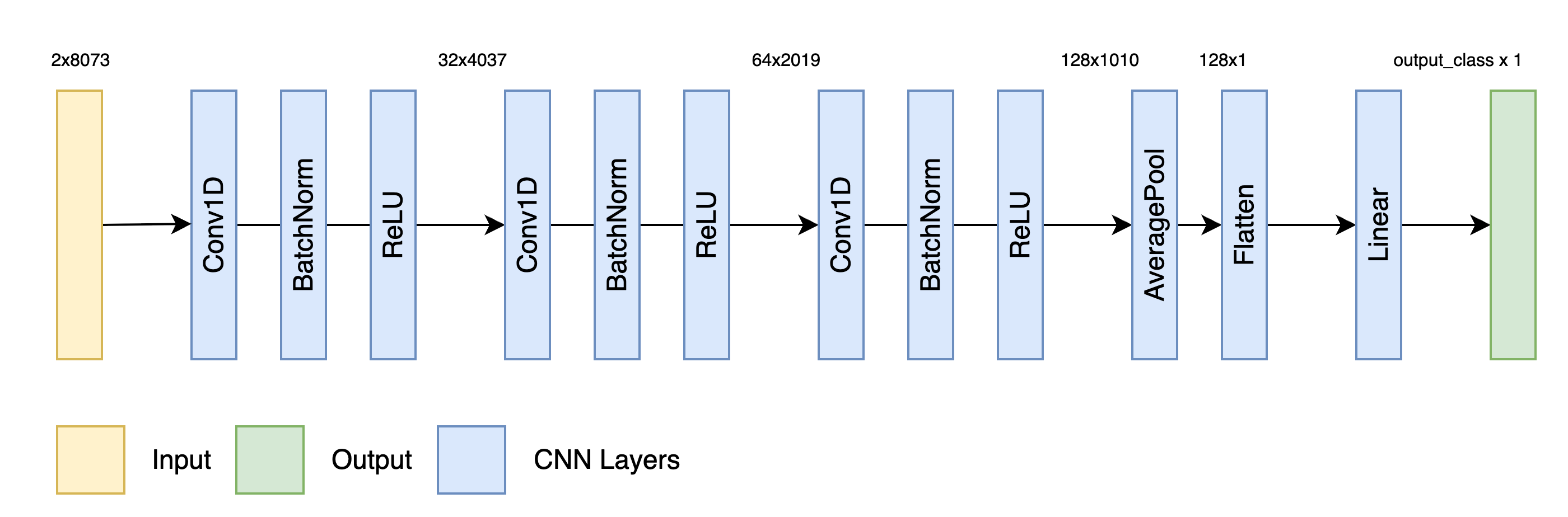}
    \caption{Architecture for the CNN Recommender}
    \label{fig:cnn_recommender}
\end{figure}

This section introduces three dedicated classifiers: noise type identification, interference type identification, and binary SIR classification implemented using a common CNN-based architecture. The network consists of three successive one-dimensional convolutional layers, each followed by batch normalization and ReLU activation. These layers progressively extract temporal and spectral features that characterize the interference waveform as per required output. The final convolutional feature maps are aggregated using an adaptive average pooling layer, which reduces the temporal dimension to a fixed-size representation regardless of the input length. A fully connected layer then projects this representation to the output classes as defined by the particular classifier. \Cref{fig:cnn_recommender} shows the architecture implemented for these CNN classifiers.

\subsection{Noise Classifier}

\begin{figure}[!htb]
    \centering
    \includegraphics[scale = 0.5]{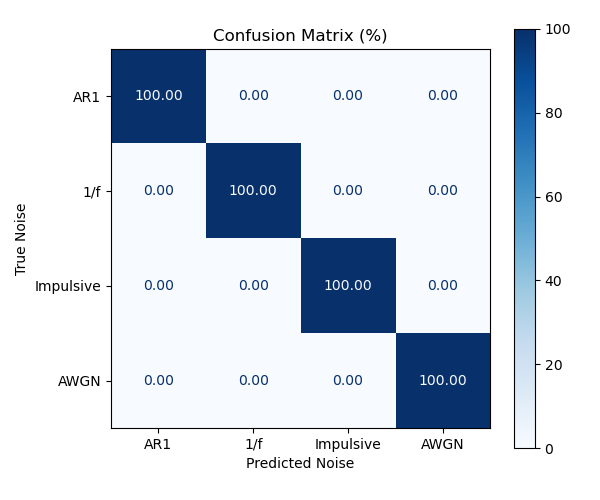}
    \caption{Confusion Matrix for Noise Classifier}
    \label{fig:noise_confusion}
\end{figure}

The first classifier, referred to as the Noise Classifier, is trained to categorize different noise environments in interference-plus-noise scenarios. Its purpose is to guide the interference mitigation pipeline by identifying the dominant noise characteristics (e.g., AWGN, AR(1), 1/f, or implusive noise), allowing the system to select the most suitable denoising model. The classifier’s performance, which can be seen in the confusion matrix in \Cref{fig:noise_confusion} shows that the classifier is able to distinctly identify each of the noise types with perfect accuracy on the testing dataset.

\subsection{Interference Classifier}
\begin{figure}[!htb]
    \centering
    \includegraphics[scale = 0.5]{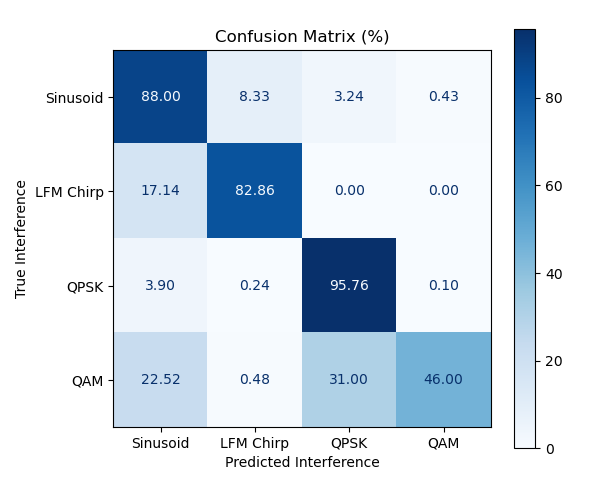}
    \caption{Confusion Matrix for Interference Classifier}
    \label{fig:interference_confusion}
\end{figure}

The second classifier, referred to as the Interference Classifier, is trained to identify the nature of interference present in signal-of-interest (SoI) plus interference plus noise conditions. This includes interference forms such as single-tone sinusoid, LFM chirp, QPSK-modulated, and QAM-modulated signals. The model’s classification accuracy across different SIR regimes is analyzed to assess robustness under challenging interference conditions. It can be seen in \Cref{fig:interference_confusion} that the interference classifier shows good performance in classifying simpler sinusoidal interference, as well as LFM chirp and QPSK interference. It performs poorly in classifying QAM interference at higher SIR's when the QPSK SoI is stronger than the QAM interference. Since the QPSK and QAM interference signals share the same bandwidth and pulse-shaping, the classifier wrongly predicts the QAM interference at higher SIR's where the overpowering signal is a QPSK signal.

\subsection{Binary SIR Classifier}

The Binary Signal-to-Interference Ratio (SIR) classifier is designed to determine whether the interference in a received signal is stronger or weaker than the desired signal. In other words, it performs a binary classification task to identify if the SIR is below or above 0 dB. This helps distinguish between interference-dominated and signal-dominated conditions.

\begin{figure}[!htb]
    \centering
    \includegraphics[scale = 0.5]{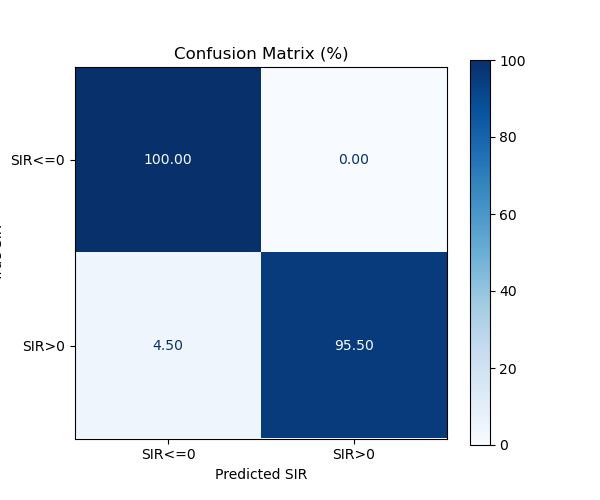}
    \caption{Confusion Matrix for Binary SIR Classifier}
    \label{fig:sir_confusion}
\end{figure}

The classifier uses the same CNN-based architecture as the noise and interference classifiers. This classifier is particularly useful for adaptive interference mitigation methods such as the Successive Interference Cancellation (SIC) algorithm. By identifying whether the interference is stronger or weaker than the signal, the system can choose the right strategy. For example, if interference dominates, it can be estimated and removed first; otherwise, the signal can be detected directly as per the system architecture discussed in \Cref{fig:sic_system}. \Cref{fig:sir_confusion} shows the near perfect classification of the SIR regions thus showcasing it's effectiveness as a preliminary step before cancellation using the SIC approach.

\subsection{End-to-End Pipeline}
\begin{figure}[!htb]
    \centering
    \includegraphics[width=1\linewidth]{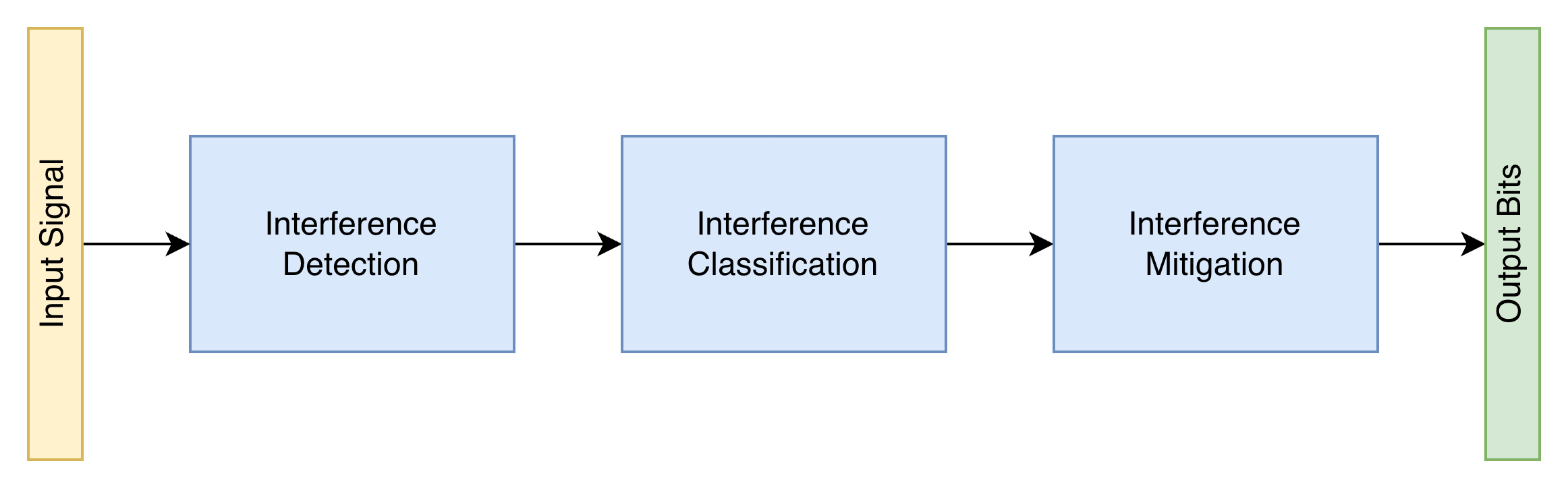}
    \caption{Proposed End-to-End Pipeline}
    \label{fig:pipeline}
\end{figure}

Finally, a multi-stage end-to-end system can be developed that integrates the outputs of the individual modules presented in the research. The first stage of this system will be the interference detection model. The interference detection block first analyzes the incoming signal to determine whether interference is present. If interference is detected, the signal is passed to the interference classification block where the interference/noise will get classified. The interference classification block consists of the aforementioned noise and interference classifiers. The output of these classifiers can be fed to the interference mitigation block which represents the different U-Net and SICU-Net systems discussed in \Cref{fig:unet,fig:sicunet}. The output of the Interference Classification block can be used to dynamically select the appropriate U-Net autoencoder models and system for the respective noise/interference types which were explored in the previous \Cref{sec:s_n,sec:s_i_n}. \Cref{fig:pipeline} shows the high-level block diagram of the proposed flow for the end-to-end system. This adaptive framework enhances overall system flexibility by ensuring that the most effective autoencoder model is applied based on the identified noise or interference setting.

\section{Conclusions}
The experimental results underscored the clear advantages of convolutional autoencoder-based approaches compared to classical methods such as matched filtering and SIC for most of the scenarios. These findings affirm the viability and effectiveness of deep learning methodologies for robust interference mitigation in modern wireless communication systems. Preliminary work on developing classifiers for the noise and interference classification also showcases the possibility of creating a system can classify and then mitigate the corresponding noise/interference using the trained autoencoder.

Although the transmitted QPSK bits and thereby symbols are randomly generated, the resulting waveform is not unstructured. The QPSK signal follows a well-defined mathematical model governed by symbol mapping and pulse shaping, which impose deterministic relationships between the in-phase and quadrature components over time. This structure allows learning-based models to extract meaningful patterns, such as symbol transitions, phase constellations, and spectral characteristics, that distinguish QPSK from noise or interference. During detection tasks, the model learns to recognize these structured signal features. While during interference mitigation, it learns to reconstruct the QPSK waveform by exploiting its consistent phase and amplitude relationships. Thus, even though the bit sequences are random, the underlying signal possesses structure that can be effectively modeled by deep neural networks.

\bibliography{IEEEabrv,./references}

\begin{thebibliography}{10}
\providecommand{\url}[1]{#1}
\csname url@samestyle\endcsname
\providecommand{\newblock}{\relax}
\providecommand{\bibinfo}[2]{#2}
\providecommand{\BIBentrySTDinterwordspacing}{\spaceskip=0pt\relax}
\providecommand{\BIBentryALTinterwordstretchfactor}{4}
\providecommand{\BIBentryALTinterwordspacing}{\spaceskip=\fontdimen2\font plus
\BIBentryALTinterwordstretchfactor\fontdimen3\font minus \fontdimen4\font\relax}
\providecommand{\BIBforeignlanguage}[2]{{%
\expandafter\ifx\csname l@#1\endcsname\relax
\typeout{** WARNING: IEEEtran.bst: No hyphenation pattern has been}%
\typeout{** loaded for the language `#1'. Using the pattern for}%
\typeout{** the default language instead.}%
\else
\language=\csname l@#1\endcsname
\fi
#2}}
\providecommand{\BIBdecl}{\relax}
\BIBdecl

\bibitem{oyedare_interference_2022}
\BIBentryALTinterwordspacing
T.~Oyedare, V.~K. Shah, D.~J. Jakubisin, and J.~H. Reed, ``Interference {Suppression} {Using} {Deep} {Learning}: {Current} {Approaches} and {Open} {Challenges},'' \emph{IEEE Access}, vol.~10, pp. 66\,238--66\,266, 2022. [Online]. Available: \url{https://ieeexplore.ieee.org/document/9802083/}
\BIBentrySTDinterwordspacing

\bibitem{lancho_rf_2024}
\BIBentryALTinterwordspacing
A.~Lancho, A.~Weiss, G.~C.~F. Lee, T.~Jayashankar, B.~Kurien, Y.~Polyanskiy, and G.~W. Wornell, ``{RF} {Challenge}: {The} {Data}-{Driven} {Radio} {Frequency} {Signal} {Separation} {Challenge},'' Sep. 2024, arXiv:2409.08839 [eess]. [Online]. Available: \url{http://arxiv.org/abs/2409.08839}
\BIBentrySTDinterwordspacing

\bibitem{almazrouei_deep_2019}
\BIBentryALTinterwordspacing
E.~Almazrouei, G.~Gianini, N.~Almoosa, and E.~Damiani, ``A {Deep} {Learning} {Approach} to {Radio} {Signal} {Denoising},'' in \emph{2019 {IEEE} {Wireless} {Communications} and {Networking} {Conference} {Workshop} ({WCNCW})}.\hskip 1em plus 0.5em minus 0.4em\relax Marrakech, Morocco: IEEE, Apr. 2019, pp. 1--8. [Online]. Available: \url{https://ieeexplore.ieee.org/document/8902756/}
\BIBentrySTDinterwordspacing

\bibitem{yapar_demucs_2024}
\BIBentryALTinterwordspacing
{\c{C}}.~Yapar, F.~Jaensch, J.~C. Hauffen, F.~Pezone, P.~Jung, S.~K. Dehkordi, and G.~Caire, ``Demucs for {Data}-{Driven} {RF} {Signal} {Denoising},'' in \emph{2024 {IEEE} {International} {Conference} on {Acoustics}, {Speech}, and {Signal} {Processing} {Workshops} ({ICASSPW})}.\hskip 1em plus 0.5em minus 0.4em\relax Seoul, Korea, Republic of: IEEE, Apr. 2024, pp. 95--96. [Online]. Available: \url{https://ieeexplore.ieee.org/document/10627485/}
\BIBentrySTDinterwordspacing

\bibitem{van_luong_deep_2022}
\BIBentryALTinterwordspacing
T.~Van~Luong, N.~Shlezinger, C.~Xu, T.~M. Hoang, Y.~C. Eldar, and L.~Hanzo, ``Deep {Learning} {Based} {Successive} {Interference} {Cancellation} for the {Non}-{Orthogonal} {Downlink},'' \emph{IEEE Transactions on Vehicular Technology}, vol.~71, no.~11, pp. 11\,876--11\,888, Nov. 2022. [Online]. Available: \url{https://ieeexplore.ieee.org/document/9837450/}
\BIBentrySTDinterwordspacing

\bibitem{yun_goh_iterative_2024}
\BIBentryALTinterwordspacing
D.~W. Yun~Goh, K.~C. Teh, S.~G. Razul, and E.~Gunawan, ``Iterative {SICNet} for {Non}-{Orthogonal} {Multiple} {Access},'' in \emph{{TENCON} 2024 - 2024 {IEEE} {Region} 10 {Conference} ({TENCON})}.\hskip 1em plus 0.5em minus 0.4em\relax Singapore, Singapore: IEEE, Dec. 2024, pp. 1873--1878. [Online]. Available: \url{https://ieeexplore.ieee.org/document/10902842/}
\BIBentrySTDinterwordspacing

\bibitem{yang_deep_2020}
\BIBentryALTinterwordspacing
Z.~Yang, C.~Yu, J.~Xiao, and B.~Zhang, ``\BIBforeignlanguage{en}{Deep residual detection of radio frequency interference for {FAST}},'' \emph{\BIBforeignlanguage{en}{Monthly Notices of the Royal Astronomical Society}}, vol. 492, no.~1, pp. 1421--1431, Feb. 2020. [Online]. Available: \url{https://academic.oup.com/mnras/article/492/1/1421/5700555}
\BIBentrySTDinterwordspacing

\bibitem{ghanney_radio_2020}
\BIBentryALTinterwordspacing
Y.~Ghanney and W.~Ajib, ``Radio {Frequency} {Interference} {Detection} using {Deep} {Learning},'' in \emph{2020 {IEEE} 91st {Vehicular} {Technology} {Conference} ({VTC2020}-{Spring})}.\hskip 1em plus 0.5em minus 0.4em\relax Antwerp, Belgium: IEEE, May 2020, pp. 1--5. [Online]. Available: \url{https://ieeexplore.ieee.org/document/9129612/}
\BIBentrySTDinterwordspacing

\bibitem{gu_radio_2024}
\BIBentryALTinterwordspacing
F.~Gu, L.~Hao, B.~Liang, S.~Feng, S.~Wei, W.~Dai, Y.~Xu, Z.~Li, and Y.~Dao, ``Radio {Frequency} {Interference} {Detection} {Using} {Efficient} {Multi}-{Scale} {Convolutional} {Attention} {UNet},'' 2024, publisher: arXiv Version Number: 1. [Online]. Available: \url{https://arxiv.org/abs/2404.00277}
\BIBentrySTDinterwordspacing

\bibitem{akeret_radio_2017}
\BIBentryALTinterwordspacing
J.~Akeret, C.~Chang, A.~Lucchi, and A.~Refregier, ``\BIBforeignlanguage{en}{Radio frequency interference mitigation using deep convolutional neural networks},'' \emph{\BIBforeignlanguage{en}{Astronomy and Computing}}, vol.~18, pp. 35--39, Jan. 2017. [Online]. Available: \url{https://linkinghub.elsevier.com/retrieve/pii/S2213133716301056}
\BIBentrySTDinterwordspacing

\bibitem{zhao_end--end_2021}
\BIBentryALTinterwordspacing
R.~Zhao, J.~Wang, and J.~Li, ``An {End}-to-{End} {Demodulation} {System} {Based} on {Convolutional} {Neural} {Networks},'' \emph{Journal of Physics: Conference Series}, vol. 2026, no.~1, p. 012006, Sep. 2021. [Online]. Available: \url{https://iopscience.iop.org/article/10.1088/1742-6596/2026/1/012006}
\BIBentrySTDinterwordspacing

\bibitem{qiang_demodulation_2021}
\BIBentryALTinterwordspacing
Z.~Qiang, T.~Yan, and C.~Tang, ``Demodulation of {Low} {SNR} {QPSK} {Signal} {Based} on {CNN},'' in \emph{2021 {International} {Conference} on {Intelligent} {Transportation}, {Big} {Data} \& {Smart} {City} ({ICITBS})}.\hskip 1em plus 0.5em minus 0.4em\relax Xi'an, China: IEEE, Mar. 2021, pp. 638--641. [Online]. Available: \url{https://ieeexplore.ieee.org/document/9525856/}
\BIBentrySTDinterwordspacing

\bibitem{shevitski_digital_2021}
\BIBentryALTinterwordspacing
B.~Shevitski, Y.~Watkins, N.~Man, and M.~Girard, ``Digital {Signal} {Processing} {Using} {Deep} {Neural} {Networks},'' 2021, version Number: 1. [Online]. Available: \url{https://arxiv.org/abs/2109.10404}
\BIBentrySTDinterwordspacing

\bibitem{ronneberger_u-net_2015}
\BIBentryALTinterwordspacing
O.~Ronneberger, P.~Fischer, and T.~Brox, ``U-{Net}: {Convolutional} {Networks} for {Biomedical} {Image} {Segmentation},'' May 2015, arXiv:1505.04597 [cs]. [Online]. Available: \url{http://arxiv.org/abs/1505.04597}
\BIBentrySTDinterwordspacing

\end{thebibliography}
\bibliographystyle{IEEEtran}

\vspace{12pt}
% \color{red}
% IEEE conference templates contain guidance text for composing and formatting conference papers. Please ensure that all template text is removed from your conference paper prior to submission to the conference. Failure to remove the template text from your paper may result in your paper not being published.

\end{document}